%% file: debris_paper_v6.tex
\documentclass[12pt,preprint]{aastex}
\usepackage{epstopdf}
\usepackage{graphicx}
\usepackage{amsmath}
\newcommand{\msun}{M$_{\odot}$\,}

\begin{document}

\title{Binaries Among Debris Disk Stars} 

\author{David R. Rodriguez\altaffilmark{1,2} and  B.\ Zuckerman\altaffilmark{1}}

\altaffiltext{1}{Dept.\ of Physics \& Astronomy, University of California, Los Angeles 90095, USA}
\altaffiltext{2}{Current address: Departamento de Astronom{\'i}a, Universidad de Chile, Casilla 36-D, Santiago, Chile (drodrigu@das.uchile.cl)}

\begin{abstract}
We have gathered a sample of 112 main-sequence stars with known debris disks.
We collected published information and performed adaptive optics observations at Lick Observatory to determine if these debris disks are associated with binary or multiple stars. 
We discovered a previously unknown M-star companion to HD~1051 at a projected separation of 628~AU.
We found that $25\pm4$\% of our debris disk systems are binary or triple star systems, substantially less than the expected $\sim50$\%.
The period distribution for these suggests a relative lack of systems with 1--100~AU separations.
Only a few systems have blackbody disk radii comparable to the binary/triple separation. 
Together, these two characteristics suggest that binaries with intermediate separations of 1--100~AU readily clear out their disks.
We find that the fractional disk luminosity, as a proxy for disk mass, is generally lower for multiple systems than for single stars at any given age.
Hence, for a binary to possess a disk (or form planets) it must either be a very widely separated binary with disk particles orbiting a single star or it must be a small separation binary with a circumbinary disk.

\end{abstract} 


\section{Introduction}\label{intro}

Planet formation occurs around young stars in disks that are rich in gas and dust, some of which can be used to form Jovian-class planets.
This needs to occur fairly rapidly, since disk gas generally dissipates over a period of a few million years \citep{Haisch:2001,Uzpen:2009}.
Eventually, even the dust in the system will be removed, either through accretion onto larger objects, stellar winds, or radiative processes. 
However, if the system has formed planetesimals or larger-sized objects, collisions can occur and produce a second generation of dust.
These dusty systems, known as debris disks, could then contain detectable quantities of dust with little or no gas present and they would be older than their gas-rich counterparts \citep[and references therein]{Zuckerman:2001,Wyatt:2008}.
The first debris disks were found by the Infrared Astronomy Satellite (IRAS).
IRAS surveyed almost the entire sky at 12, 25, 60, and 100\micron~and discovered infrared excesses around many stars including Vega, Fomalhaut, $\beta$~Pictoris, and $\epsilon$~Eridani.
The solar system's own Kuiper Belt may be analogous to these circumstellar disks \citep[e.g.,][and references therein]{Luu:2002}.
A complete understanding of the formation and evolution of planetary systems requires knowledge of the properties of disks, from the gas-rich protoplanetary disks to the gas-poor debris disks.

Over 500 extrasolar planets have been discovered so far. Most extrasolar planet searches have utilized the precision radial velocity technique; however, the Kepler satellite has recently reported over 1200 candidate planets (see \citealt{Borucki:2011}). While highly successful, the radial velocity technique is generally applied to single stars or very widely separated binaries. 
The problem is that the spectra of close binary stars are highly variable due to their orbital motion which typically leads to relatively large velocity uncertainties.
However, the TATOOINE radial velocity survey \citep{Konacki:2009} is searching for planets around double-lined spectroscopic binaries and has reached precisions of a few m/s, comparable to, albeit not as good as, precisions reached for single stars. Although \citet{Konacki:2009} have yet to report a circumbinary planet and their sample size is small (10 systems), their work demonstrates that future planet searches may be performed on close binaries.

\citet{Duquennoy:1991} estimate that 57\% of G stars may be in multiple systems.
Other, more recent surveys show single stars are somewhat more common, but the fraction of multiples is still about $\sim50$\% \citep{Eggleton:2008,Raghavan:2010}.
A question naturally arises: if so many stars are in multiple systems, what does this say about the formation and evolution of disks and planets?
About 20\% of known extra solar planets reside in wide 
separation binaries \citep{Raghavan:2006,Eggenberger:2007},
most of which have separations of 100s of AU.
%No unambiguous detection of an extrasolar planet has been made for a binary with separation $<$20~AU, though eclipse timing variations of HW~Virginis, CM~Draconis, and NN~Serpentis (a post-common envelope binary) suggest that planets may orbit these binaries \citep{Deeg:2008,Lee:2009,Beuermann:2010}.
Eclipse timing variations of HW~Virginis, CM~Draconis, and NN~Serpentis (a post-common envelope binary) suggest that planets may orbit these binaries \citep{Deeg:2008,Lee:2009,Beuermann:2010}.
Despite these efforts (see also \citealt{Konacki:2009}), detection of planets in circumbinary orbits remains a challenge. 
While this manuscript was being prepared, \citet{Doyle:2011} presented the discovery of Kepler-16b, the first circumbinary planet among the Kepler data. Both stars and the planet share a common orbital plane, suggesting the planet formed in a circumbinary disk.
Indirect evidence implies the existence of rocky planets orbiting the close (3.4-day) main sequence binary BD +20~307. This system displays a large quantity of warm dust in the terrestrial planet zone \citep{Song:2005} that likely is the aftermath of a collision of two rocky planets that orbit this $\sim$1~Gyr old binary system \citep{Zuckerman:2008b, Weinberger:2011}.
The study of circumstellar and circumbinary disks, then, can be used to comment on the process of planet formation around binary stars.

There has been some previous effort to address the issue of dusty disks in binary systems. 
Sub-millimeter studies of young ($\lesssim5$~Myr) binaries have shown that binary stars with intermediate separations ($1<a<50-100$~AU) have lower sub-millimeter fluxes than more widely separated binaries or single stars (\citealt{Jensen:1996}).
Interferometric observations of nearby $\sim$8~Myr-old disks in a triple and quadruple system have shown the disks to be truncated by the nearby stellar companions \citep{Andrews:2010}, as expected from numerical simulations \citep{Lubow:2000,Artymowicz:1994}.
Among pre-main sequence stars, several studies have found the disk lifetimes of small to moderate separation binaries to be shorter than that of single stars or very widely separated binaries \citep{Bouwman:2006,Cieza:2009}.
Recently, \citet{Kraus:2011} presented results in the Taurus-Auriga star forming region demonstrating that binaries with separations $\lesssim$40~AU readily disperse protoplanetary disks.
\cite{Trilling:2007} used the MIPS camera on the Spitzer Space Telescope to search for infrared excess among 69 known binaries. They not only found that some binary systems have debris disks, but also that that the incidence of debris disks among binaries is marginally higher than for single AFGK stars older than 600 Myr. 

\section{Sample} \label{debris:sample}

We have approached the question of stellar multiplicity among debris disk systems from a different direction than did \cite{Trilling:2007}.
Whereas they began with a asample of known binarity, we instead selected a sample of stars with known infrared excesses that satisfy two additional criteria criteria: ages older than 10 million years, to reduce the chances that we include protoplanetary disks, and distances within 100 parsecs, to ensure we have sufficient information. 
We constructed our sample from \cite{Rhee:2007}, \cite{Rebull:2008}, \cite{Chen:2005}, and included BD+20 307 from \cite{Song:2005}.
We note that \citet{Mawet:2011} recently demonstrated that $\epsilon$~Cephei (HIP~109857) does not contain a debris disk and is thus not included in our sample.

Many Spitzer studies tend to be biased against binaries, so we were careful to select samples that were not biased in favor or against binaries; the \cite{Rebull:2008} study was based on stars in the $\beta$~Pictoris moving group, while \cite{Chen:2005} searched for debris disks among nearby, young (12--600~Myr) stars.
For the 60\micron\, IRAS sample from \cite{Rhee:2007}, we did not include objects marked as new candidate debris disks (see Note 2 in their Table~2).
By drawing from papers that use the whole sky with IRAS or in moving groups with Spitzer, we expect to avoid any significant bias in multiplicity fraction.

From these references we obtain stellar properties such as spectral type, age, fractional infrared luminosity ($L_{IR}/L_*$), dust temperature, and dust orbital semi-major axis. The dust properties were derived in the respective references from blackbody fits to the excess emission after modeling the stellar photosphere (however, see Section~\ref{dust_T}).
We note that the infrared excesses for the sample in \citet{Trilling:2007} were often faint, unlike those in our sample.

\begin{figure}[htb!]
\begin{center}
\includegraphics[width=14cm,angle=0]{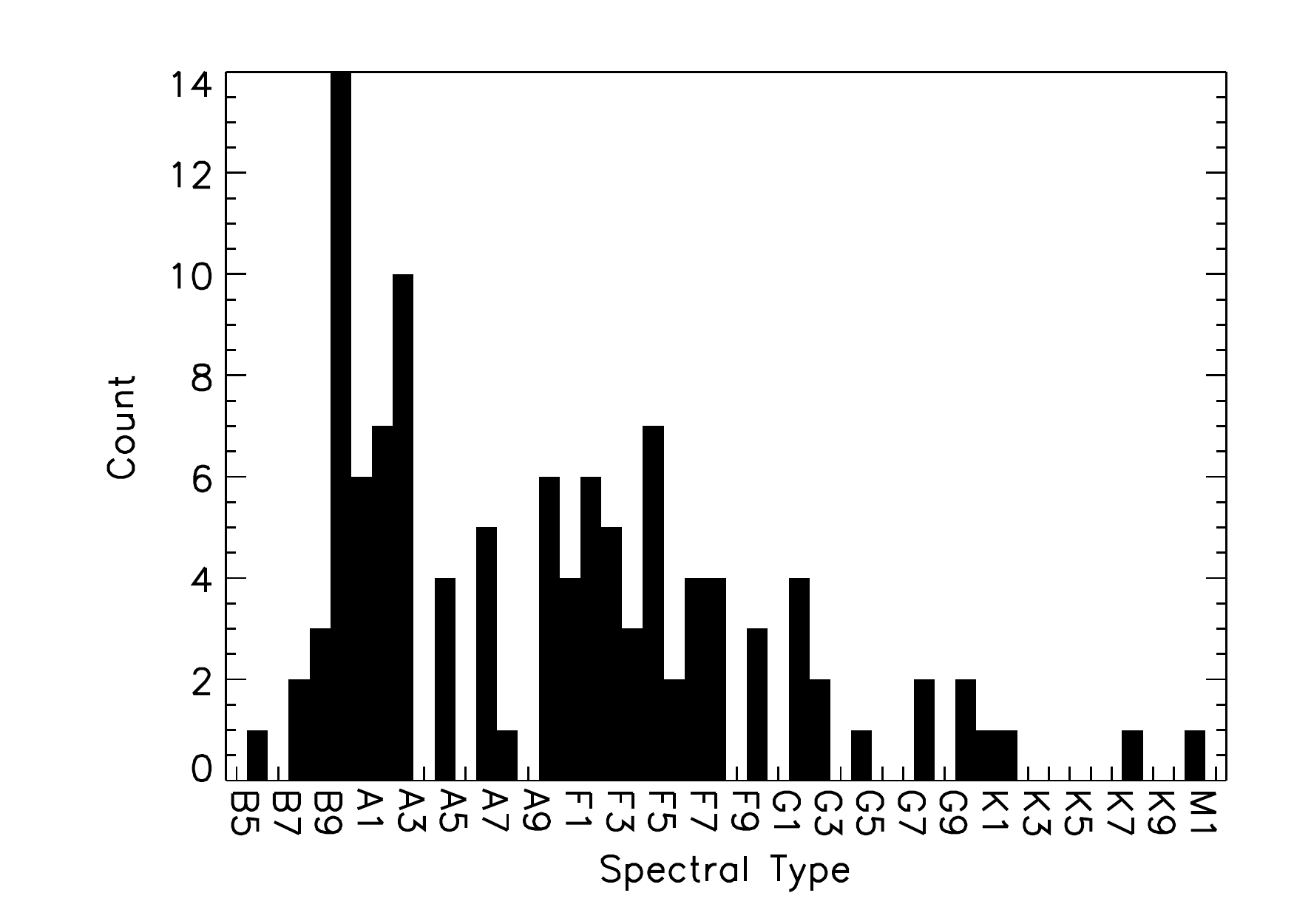}
\end{center}
\caption{Spectral type distribution of our 112-star debris disk sample.}
\label{fig:spectraltypes}
\end{figure}

Our efforts resulted in a catalog of 112 systems with spectral types essentially in the range B8 to K2, though most are A and F-type stars (see Figure~\ref{fig:spectraltypes}).
Our sample, and the gathered information, is listed in Table~\ref{tab:debris_sample}.
The majority of these stars are IRAS detections from \cite{Rhee:2007}. Many, though, have been confirmed by Spitzer. Two of our objects overlap with the \cite{Trilling:2007} sample: HIP~15197 and HIP~66704.
Table~\ref{tab:debris_sample} stars have distances ranging from 3 ($\epsilon$~Eri) to 100~pc with a median distance of 38~pc.
%an average distance of 45~pc.

\section{Procedure} \label{procedure}

After specifying the sample, we searched the literature to determine which stars are known to be multiples. 
We used a variety of catalogs to search for information, including the Ninth Catalogue of Spectroscopic Binary Orbits \citep{Pourbaix:2004}, the Hipparcos and Tycho Double Star catalog \citep{ESA:1997}, the Sixth Catalog of Orbits of Visual Binary Stars\footnote{Available at \url{http://ad.usno.navy.mil/wds/orb6.html}}, the Washington Double Star catalog \citep{Mason:2001}, the Catalog of Components of Double \& Multiple stars \citep{Dommanget:2002}, the Multiple Star Catalog \citep{Tokovinin:1997}, as well as checked SIMBAD and VizieR for papers on individual systems.
While several systems have widely-separated candidate companions, most can be readily confirmed or ruled out by examining proper motions and estimated distances.
Including our work in Section~\ref{AOSearch}, we found 28 binary or triple star systems within the sample and list these multiples in Table~\ref{tab:multiples}. 
This corresponds to a multiplicity of $25\pm4\%$ (independent of whether they are binaries or triples), where the 4\% error, as well as the uncertainties given in Table~\ref{debris:binfraction}, are estimated as described by \cite{Burgasser:2003}.
We have broken down our multiplicity fraction by spectral type in Table~\ref{debris:binfraction}, which also includes the multiple fractions obtained by \citet{Eggleton:2008} and \cite{Duquennoy:1991}.
Our triples are hierarchical, with a close pair orbited by a more distant star.

\begin{figure}[htb]
\begin{center}
\includegraphics[width=12cm,angle=0]{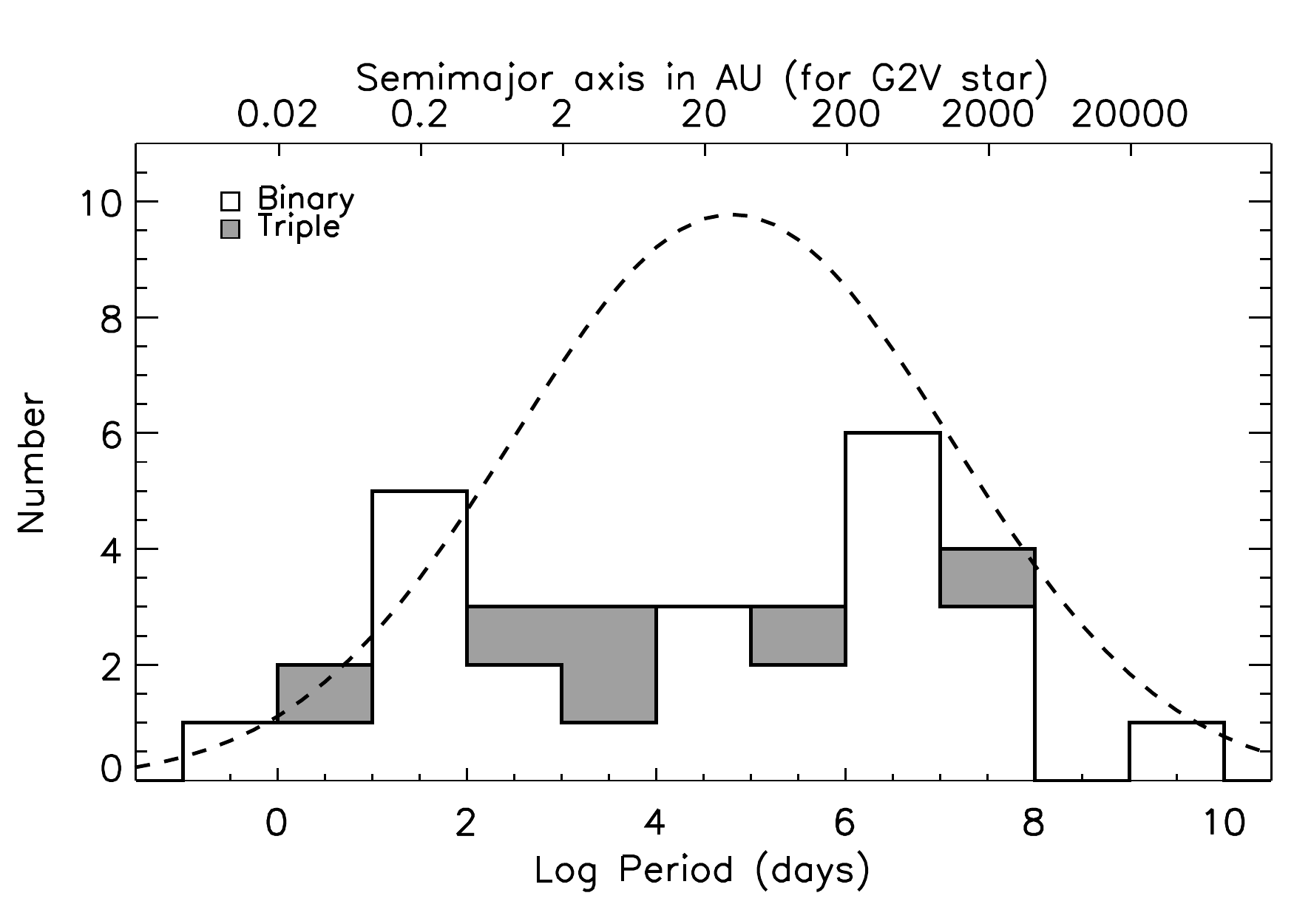}
\end{center}
\caption{Period distribution of our debris disk sample. 
%The orange represents binaries while the blue is for multiples (triples in all cases). 
For triple systems, we include both periods (for example, the A-B period and the AB-C period). 
For comparison, the semi-major axis for a sun-like star is displayed on top. The dashed line illustrates the expected period distribution from \citet{Duquennoy:1991} (and very simlar to \citealt{Eggleton:2008}) normalized to contain the expected number of binaries or multiples in the sample ($112\times0.5=56$ multiples).
While the short and long period binary distribution approximately matches the  \citet{Duquennoy:1991} distribution, there is a lack of systems with intermediate periods (separations $\sim1-100$~AU).}
\label{fig:debrisperiod}
%\epsscale{0.4}
\end{figure}

We collected period and/or semi-major axis information for our multiples.
Table~\ref{tab:multiples} lists our separations for the binary and multiple stars.
For those without periods, we estimate the period assuming the projected separation is the semi-major axis and the orbits are circular. 
The mass of the primary is estimated from its spectral type and age.
We plot all the available, or derived, periods as a histogram in Figure~\ref{fig:debrisperiod}.
The dashed line represents the period distribution of \citet{Duquennoy:1991} normalized to contain about 56 systems, the expected number of multiples if the multiplicity fraction were 50\%. 
This dashed line differs from our debris disk sample, which suggests a lack of systems with periods of about $10^2 - 10^6$ days or semi-major axes about 1--100~AU for sun-like stars.

As mentioned in Section~\ref{intro}, \citet{Jensen:1996} found that young ($\lesssim5$~Myr) binary stars with separations $1<a<50-100$~AU have lower sub-millimeter fluxes than more widely separated binary stars or single stars; this is approximately the same separation range as the gap in our sample and suggests that at such separations binary companions are effective at disrupting the formation of disks or accelerate the clearing of dust.
However, it could also be that our sample is missing 20--30 binaries or multiples with intermediate-size semi-major axes. 
While most of our sample stars are well documented in the literature, there are some with very little information of the sort that can be used to determine their binary nature.
We describe in the following section a search for missing companions. 

\section{Adaptive Optics Search for Companions}\label{AOSearch}

Wide separation companions would already have been detected with data from the Hipparcos satellite or proper motion surveys.
To detect closer companions, one can use high-resolution spectroscopy over several epochs to see if the radial motion of the star changes indicating the presence of a massive companion.
However, as most of our stars with little information are A-type stars, precise radial velocities are difficult to measure due to a relative lack of lines and rotationally-broadened profiles.
An alternative way to search for companions is to use adaptive optics (AO).
IRCAL, the infrared camera for adaptive optics at Lick observatory, has field of view of 20\arcsec\, and thus can detect companions out to nearly 1000~AU at 100~pc. 
With good seeing, companions as close as $\sim$0\farcs15 can be imaged.
Because the peak of the binary distribution is expected to be between 10--100~AU \citep{Duquennoy:1991,Eggleton:2008} and our systems are located within 100~pc, IRCAL will be capable of finding these missing companions, should they exist.
Roughly equal mass companions can be very quickly ruled out or confirmed after exposing for only seconds in the raw data alone. 
Fainter companions can be detected by co-adding dithered exposures.

We had three AO runs at Lick with IRCAL (June, October 2009; August 2010). Additional runs were scheduled in April 2010 and March 2011, but these suffered from bad weather (rain/snow).
These runs were primarily intended for observation of Herschel DEBRIS targets. 
DEBRIS, or Disc Emission via a Bias-free Reconnaissance in the Infrared/Submillimeter, is a Herschel key project of which we are a part (see \citealt{Matthews:2010}).
The DEBRIS project is observing 446 stars at 100 and 160-$\mu$m to look for far-infrared excesses indicative of cool dust in nearby star systems.
There is some overlap between the DEBRIS sample and our own sample of 112 dusty stars and it has not been difficult to observe some additional Table~\ref{tab:debris_sample} when the opportunity arises. 
Table~\ref{tab:ircalobs} lists our IRCAL AO observations of 20 Table~\ref{tab:debris_sample} debris disk systems.
Observations were carried out using the Ks and BrG-2.16 filters. If a candidate was suspected, observations at other wavelengths (J, H, Fe~II) were performed.
BrG-2.16 and Fe~II are narrow band filters centered on 2.167 and 1.644\micron, respectively. They are similar to the Ks and H filters, but $\sim10$ times narrower.

Data reduction for these observations was carried out in the usual manner (flat fielding, dark subtraction, sky subtraction) using standard IRAF routines.
Because the X and Y platescale are not the same, we used IDL routines to rescale the image while preserving flux to 76 mas/pixel in both X and Y though we later calibrate the plate scale using known binaries. 
In our search of these 20 systems we detected only 2 companions. 
One of these, HIP~35550 is a known triple system and was observed as part of the DEBRIS study.
The other, HD~1051, we describe below.

\subsection{HD~1051AB System}\label{hip1185}

HD~1051 (HIP~1185) is a 600~Myr-old A7 star located 88.3~pc away.
An infrared excess at 60 and 100\micron\, was detected around this star suggesting a debris disk with $T_\text{dust}=40$~K and $R_\text{dust}=173$~AU \citep{Rhee:2007}.
This system was observed as part of our IRCAL AO observations on 2010 August 4 with the Ks filter for a total integration time of 20s (10 frames, each being 10 coadds of 0.2s exposures).
A candidate companion, visible in the individual frames, was detected.
%A companion was detected as part of our these observations and is visible in the individual frames. 

To determine the projected separation for detected candidate companions,
we calibrated our position angle (PA) offset and plate scale using calibrator stars from the the Sixth Catalog of Orbits of Visual Binary Stars. Some of these stars were targets in the DEBRIS project, but we used them here for calibration.
We list these stars, and the results, in Table~\ref{tab:wdscalib}.
We measured $\Delta$X and $\Delta$Y, the difference in X and Y between the primary and the companion, for these stars using tasks in IRAF's daophot package and performed a least-squares fit  to:
\begin{align*}
\text{R.A.} = A \Delta X \cos \theta  - B \Delta Y \sin \theta  \\
\text{Decl.} = A \Delta X \sin \theta  + B \Delta Y \cos \theta 
\end{align*}
for $A$, $B$, and $\theta$. $A$, $B$ are the X, Y plate scales, respectively, and $\theta$ corresponds to the offset in position angle relative to the y-axis. The IRCAL detector is aligned such that the y-axis on image frames is approximately North, so $\theta$ amounts to a rotation.
Our best fit values are: $A = 72.6$~mas/pix, $B = 77.5$~mas/pix, and $\theta = 0.52$~degrees.
The rms differences in the expected position angle and separation ($\rho$) compared to the calculated values from our fit are adopted as our calibration uncertainties. The uncertainties in separation and position angle are 52~mas and 0.9~deg, respectively.
These values are accurate only for the 2010 August observing run.

\begin{figure}[htb]
\begin{center}
\includegraphics[width=12cm,angle=0]{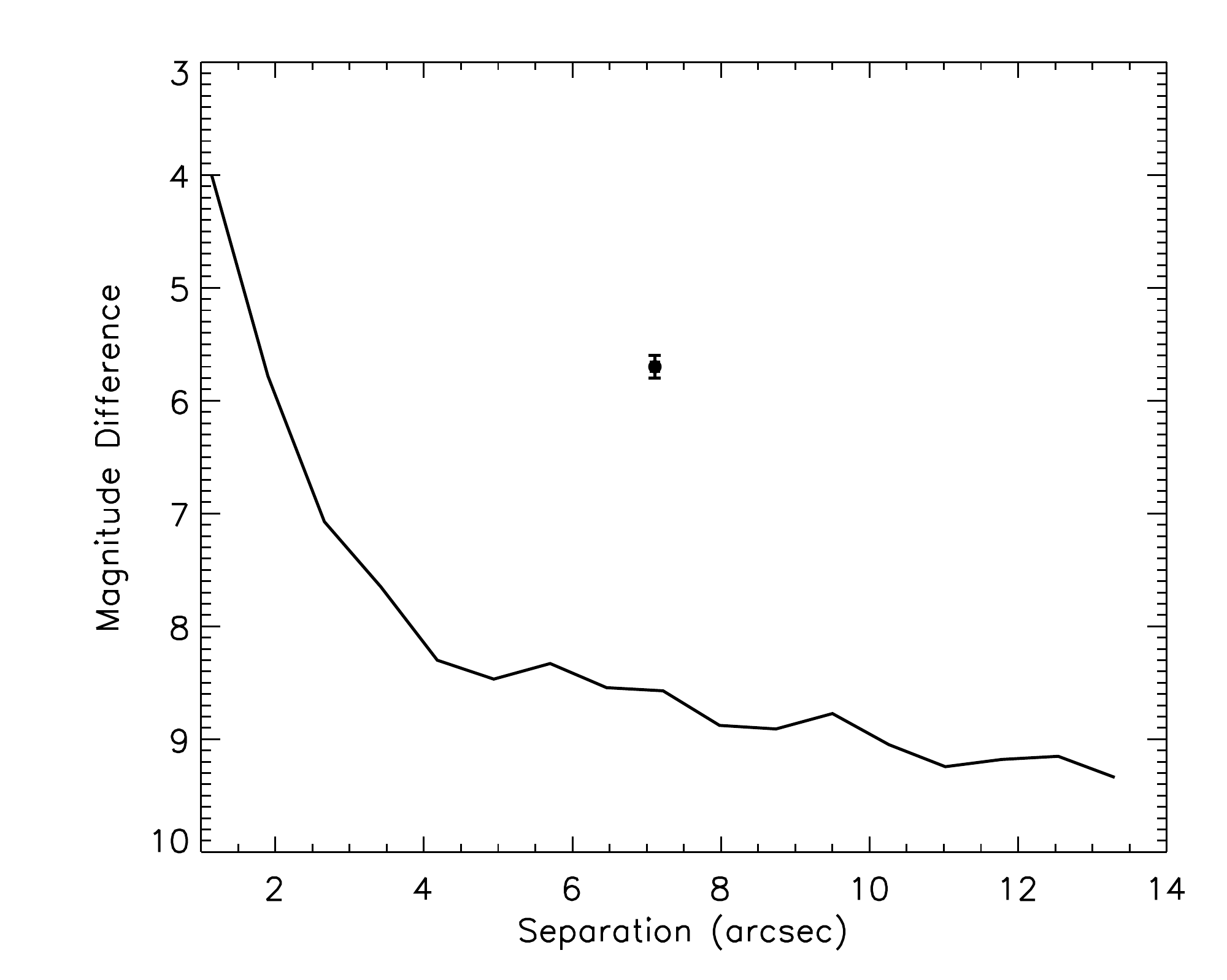}
\end{center}
\caption{Magnitude limits ($5\sigma$) in the Ks filter as a function of separation for our HD~1051 (HIP~1185) data. Limits are obtained by performing 2-pixel radius aperture photometry (the same as for HD~1051B) at random locations in annuli around the primary. The standard deviation of these measurements provides an estimate of the noise.
The detected companion is plotted as a circle.}
\label{fig:ircallimit}
%\epsscale{0.4}
\end{figure}

We measure the separation of HD~1051's companion to be $7.11\pm0.05\arcsec$
at $311.19\pm0.18$~degrees East of North. Uncertainties are standard deviations of the measurements performed on the individual frames and do not include the calibration uncertainties. 
At a distance of 88.3~pc, 7.11\arcsec\, corresponds to a projected separation of 628~AU.
Circular aperture photometry was likewise performed on the individual frames with a 2-pixel radius (0\farcs15) as this aperture maximized our signal-to-noise. We measure an apparent Ks magnitude difference of $5.7\pm0.1$.
At such separations, we could have detected companions up to 8 magnitudes fainter than the primary at the $5\sigma$ level, corresponding to $\sim0.1$\msun; see Figure~\ref{fig:ircallimit}.
The primary has an apparent 2MASS Ks magnitude of $6.25\pm0.02$, which implies the secondary has apparent magnitude of $12.0\pm0.1$, or an absolute magnitude of $7.3\pm0.1$.
The system has an estimated age of 600~Myr \citep{Rhee:2007}, which implies, when comparing to \citet{Baraffe:1998} models, that the secondary has spectral type $\sim$M3 and mass $\sim0.3$\msun.
%Its visual magnitude would then be $\sim$16.6 and thus not visible with Hipparcos.

\begin{figure}[htb]
\begin{center}
\includegraphics[width=6cm,angle=0]{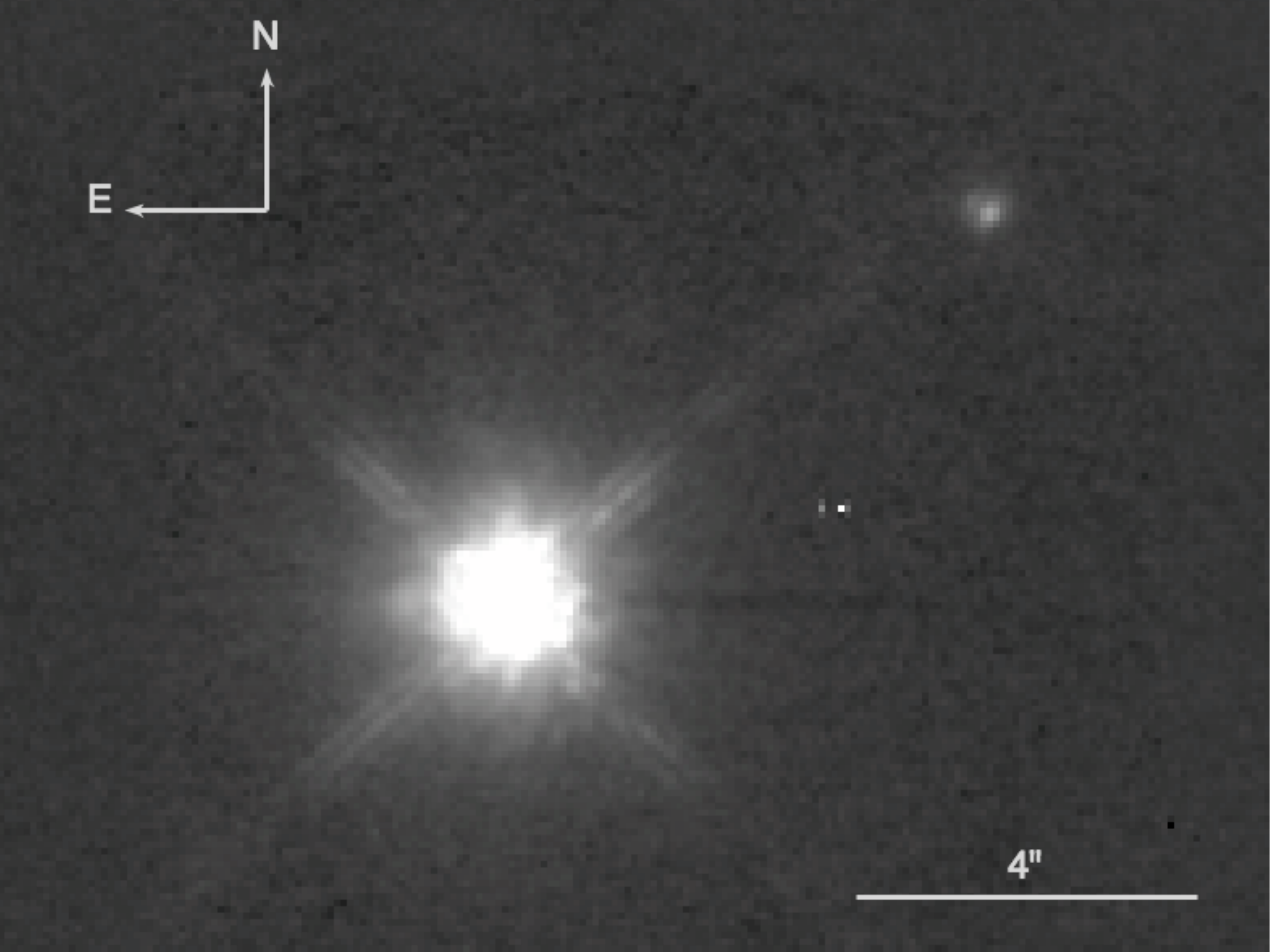}
\includegraphics[width=6cm,angle=0]{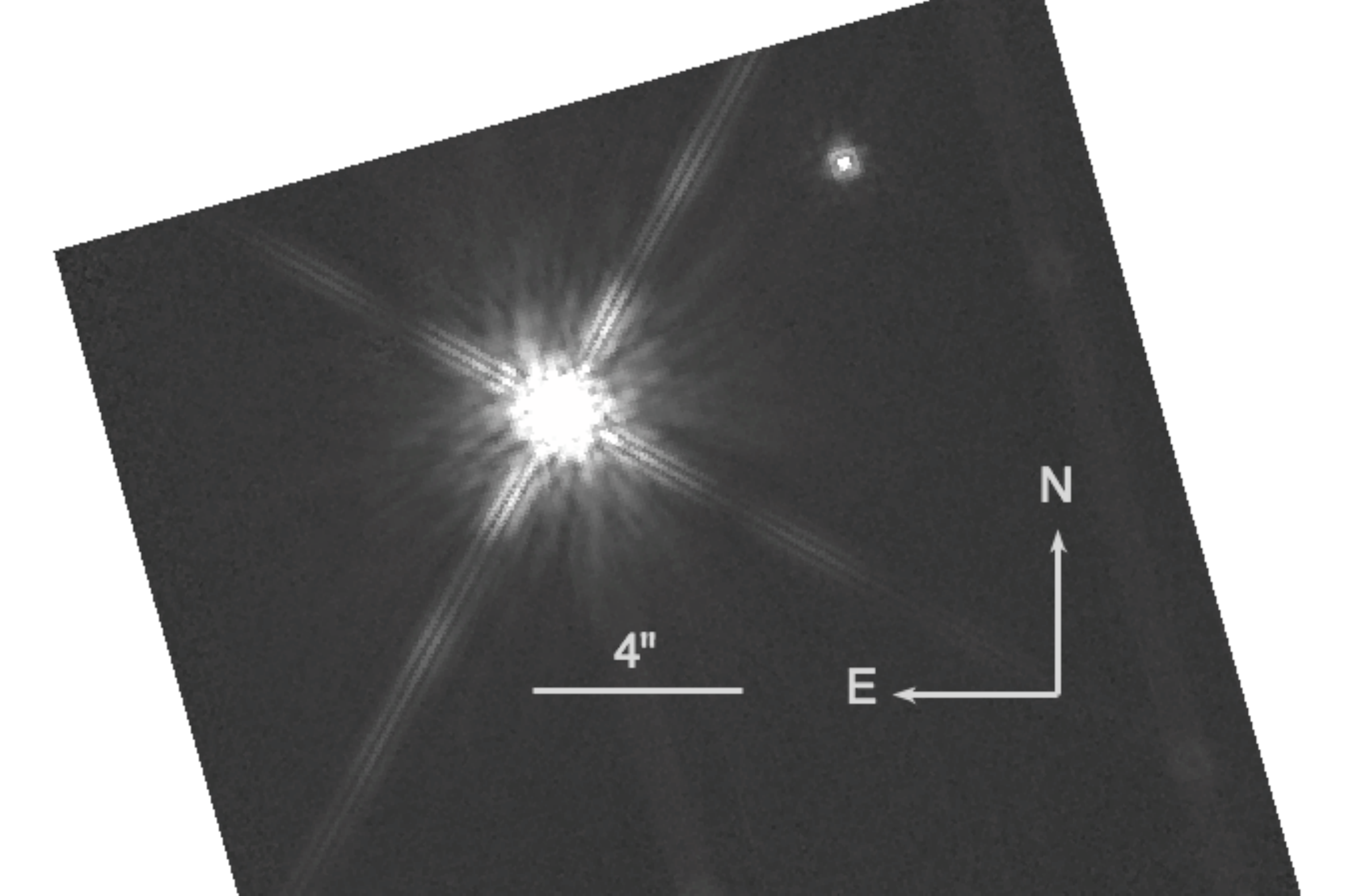}
\end{center}
\caption{Lick IRCAL (left) and HST NICMOS image (right) for HD~1051.}
\label{fig:hip1185}
\end{figure}

\begin{figure}[htb]
\begin{center}
\includegraphics[width=12cm,angle=0]{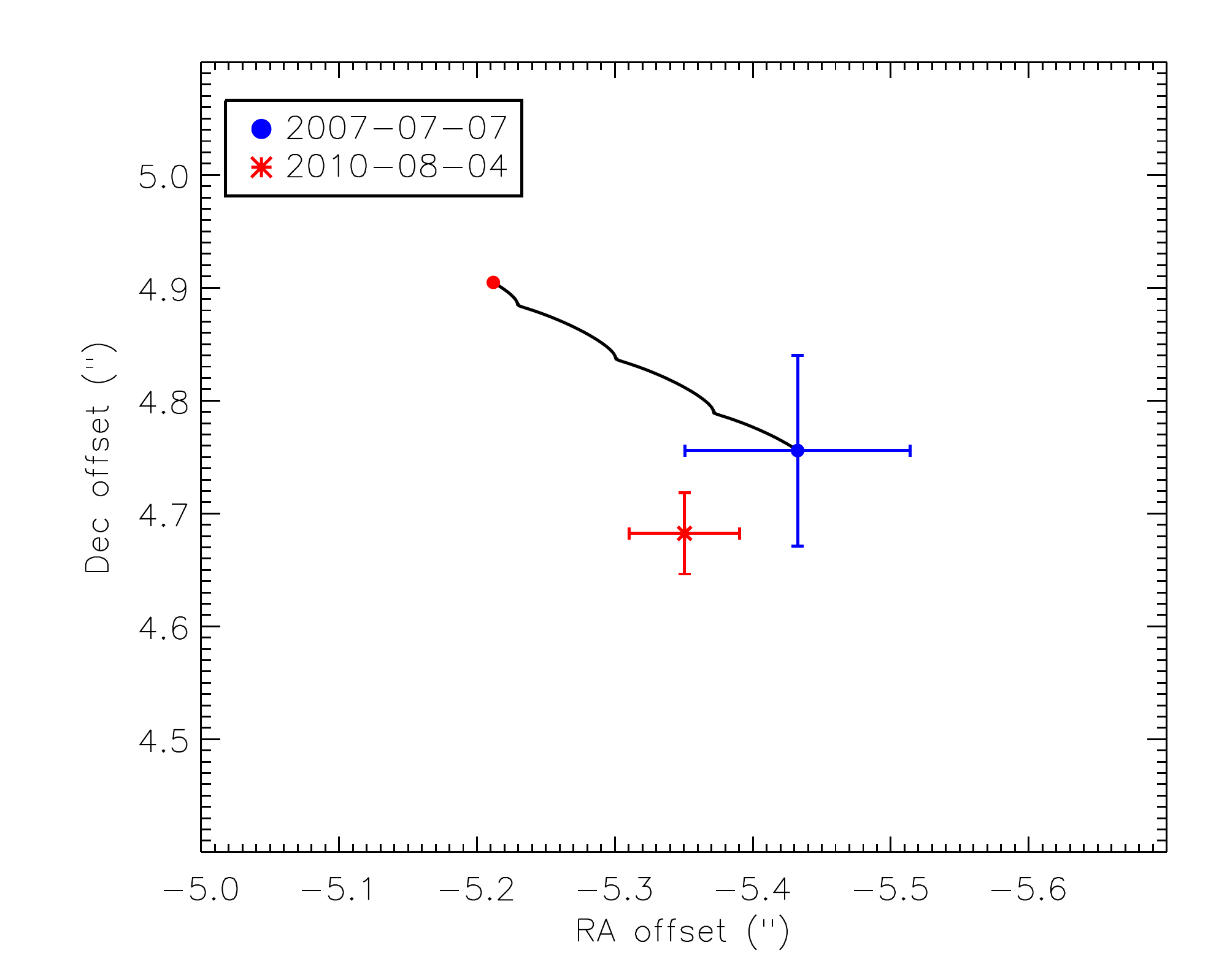}
\end{center}
\caption{Measured offsets for HD~1051B in Lick IRCAL and HST NICMOS data. The black line illustrates the motion relative to HD~1051 of a stationary background object over the 3-year period. The circle at the end of the black line is the predicted location of a background object in 2010.}
\label{fig:hip1185_move}
%\epsscale{0.4}
\end{figure}

In order to confirm the detected object is bound to the system, and thus a true companion, we compare the IRCAL 2010 data with existing HST NICMOS data taken during 2007 as part of program 11157 (PI: Joseph Rhee).
The HST location is $7.22\pm0.09\arcsec$ at $311.23\pm0.57$ degrees. As before, uncertainties are the standard deviations of multiple measurements and do not include systematics: $\sim0.08\arcsec$ in both R.A.\ and Decl.\ due to the primary's location behind the NICMOS coronographic spot.
The IRCAL and HST data are displayed in Figure~\ref{fig:hip1185} and both show the companion at approximately the same location.
Using the known distance and proper motion of the primary, as well as the measured location of the companion, we find that the secondary has not moved relative to the primary over the 3-year baseline. Figure~\ref{fig:hip1185_move} illustrates this by showing the measured location of the IRCAL and HST data and the expected motion a stationary background object would have had relative to the primary over this time period. 

\section{Dust Temperatures and Fractional Luminosities} \label{dust_T}

Dust temperatures, and hence disk radii or semi-major axes, are estimated in the various references by fitting blackbodies to the observed infrared excess.
This assumes the grains are not much smaller than the wavelengths at which they principally emit.
For objects with a detected excess only at $60\micron$, \citet{Rhee:2007} assigned a temperature of 85~K corresponding to the peak of a blackbody at that wavelength in $F_\nu$.
There are about $\sim$20 such objects in our sample.
For objects with MIPS-measured excess emission only at 70$\micron$, \citet{Rebull:2008} set a temperature of 41~K corresponding to the peak for $\lambda F_\lambda$ while \citet{Chen:2005} use a temperature of 40~K, based on a modified blackbody fit to the dust around AU~Mic.
As most of our data come from \citet{Rhee:2007}, we have re-fit the spectral energy distribution (SED) for those systems with only $70\micron$ excesses (J.\ Rhee, 2011, private communication).
The modified objects are HIP~2072, HIP~25486, HIP~66704, and HIP~92680.
In comparison, \citet{Trilling:2007} determine upper limits on temperature by using the MIPS 3$\sigma$ upper limit on the $24\micron$ emission for those objects with only $70\micron$ excesses.

\begin{figure}[htb]
\begin{center}
\includegraphics[width=10cm,angle=0]{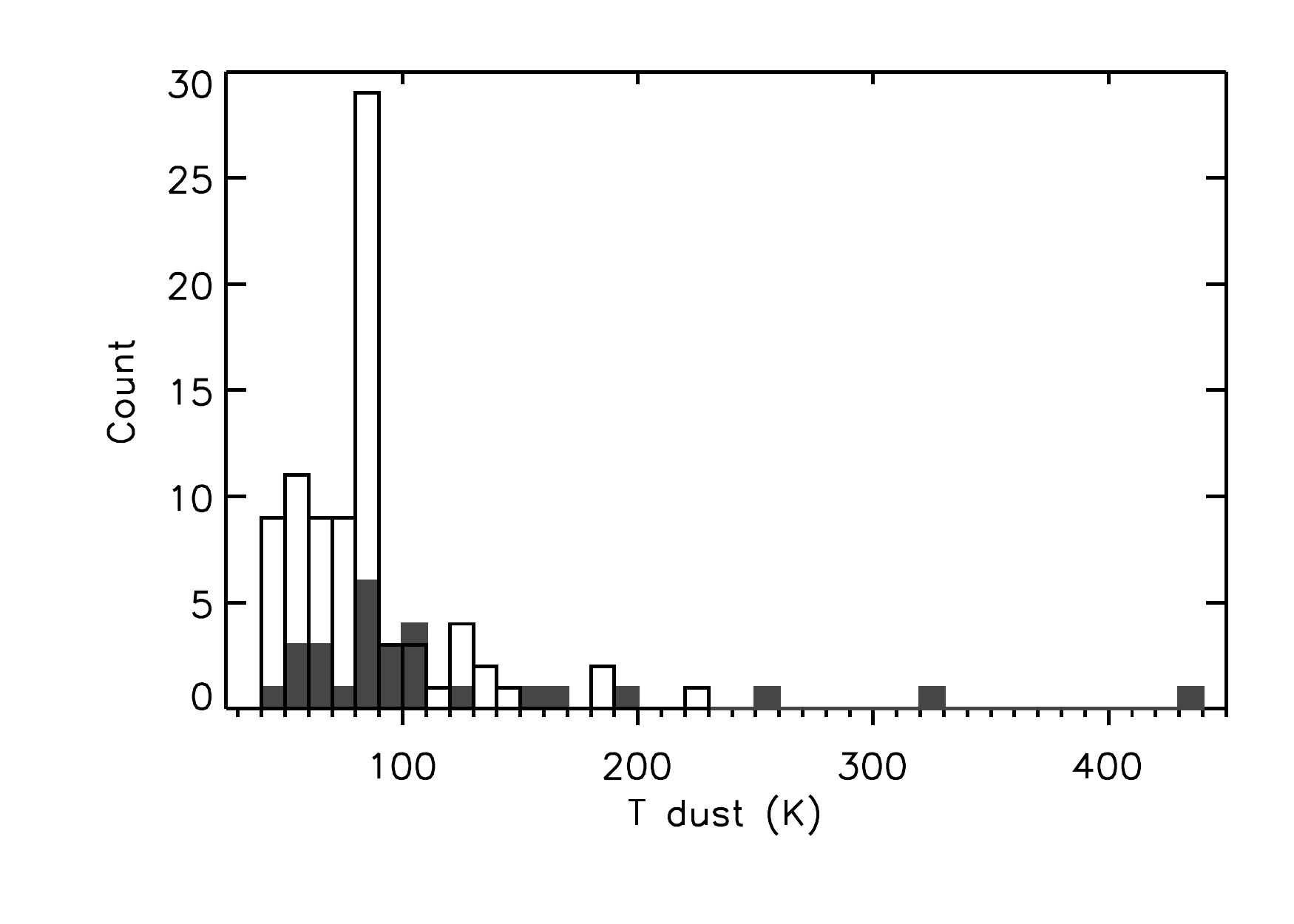}
\includegraphics[width=10cm,angle=0]{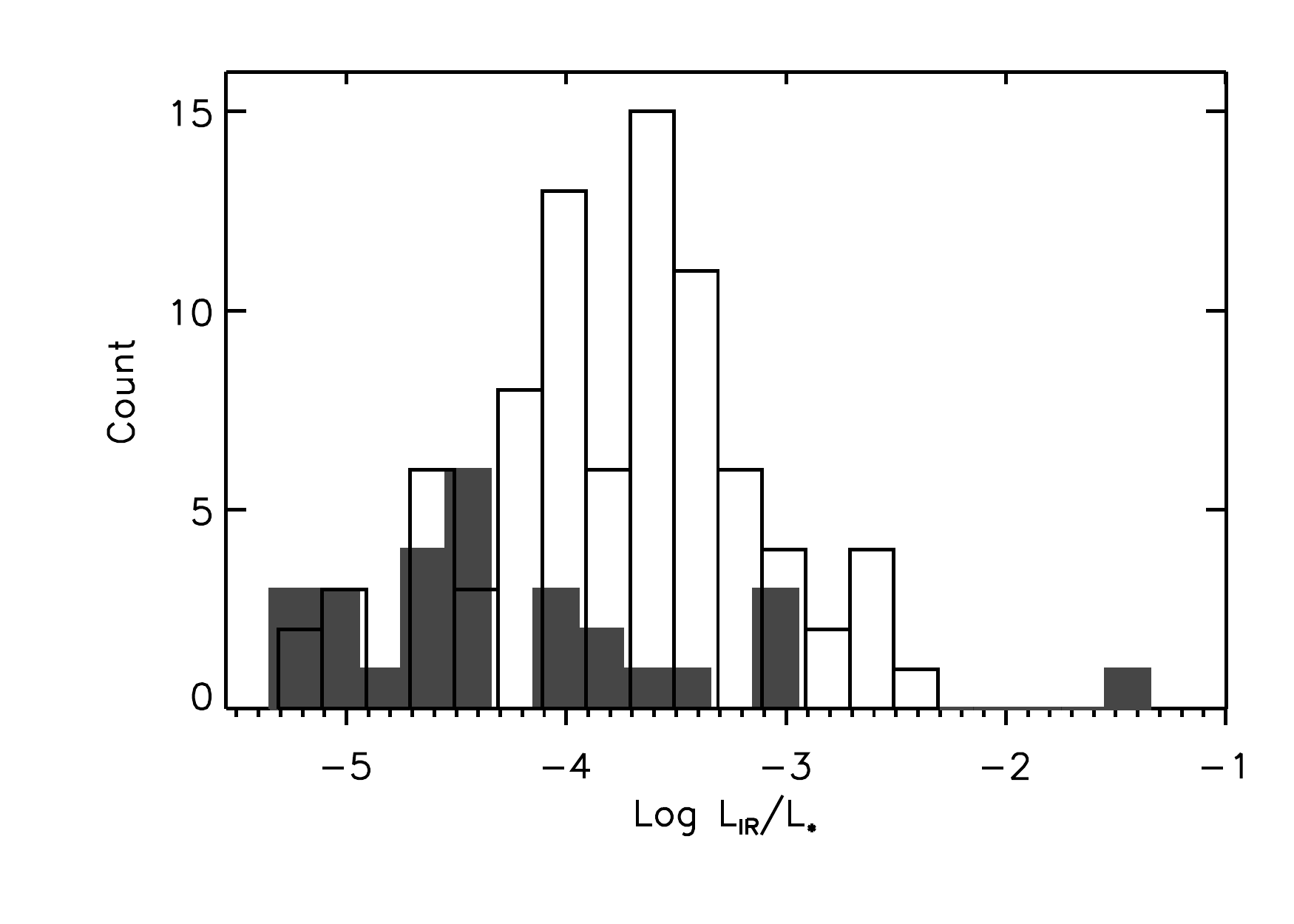}
\end{center}
\caption{
Dust temperatures (top) and fractional luminosities (bottom) for our debris disk sample.
Open bars represent single systems, while the shaded bars represent multiples.
The large number of systems in the $\sim70-90$~K bin is partially an artifact of the analysis and includes systems with detections at only 60 or 70\micron, see discussion in Section~\ref{dust_T}.
While the dust temperatures are similar for both single and multiple systems, the fractional luminosities for single stars are, on average, 2--3 times larger than those for multiple stars.
BD~+20~307 is the rightmost system in both panels \citep{Weinberger:2011}.
}
\label{fig:debrishisto}
\end{figure}

For blackbody-like grains, we estimate disk semi-major axes from the stellar radius and temperature and the dust temperature: 
\begin{align*} 
   R_\text{dust} = \frac{R_*}{2} \left(\frac{T_*}{T_\text{dust}}\right)^2 \\
\end{align*}
%Note that by fitting a blackbody that peaks at the single wavelength excess detection, one adopts the maximum possible temperature for large blackbody grains.
%Cooler blackbody temperatures, and thus larger semi-major axes, can still be consistent with single measurements at 60 or 70\micron.
The distributions of dust temperature and fractional luminosity are shown in Figure~\ref{fig:debrishisto}. While the dust temperatures are similar for both single and multiple systems, the fractional luminosities for single stars are, on average, 2--3 times larger than those for multiple stars.
%The Kolmogorov-Smirnov (KS) test can be used to determine if two samples are drawn from the same distribution. It computes an empirical distribution function and returns a statistic, $P$, which is the probability that both samples are drawn from the same underlying distribution (whatever that may be). If $P$ is small (in general smaller than about 0.05, or even 0.01 to be more conservative) then the two samples are considered to be drawn from different distributions.
We applied the Kolmogorov-Smirnov (KS) test for our sample of dust temperatures and fractional luminosities.
For the dust temperature, we find $P=0.04$, which, while small, may not allow us to rule out that they are drawn from the same distribution.
However, for fractional luminosity we find $P=2\times10^{-4}$ suggesting that the distribution of fractional luminosity between single and multiple stars is different. 

\begin{figure}[htb]
\begin{center}
\includegraphics[width=10cm,angle=0]{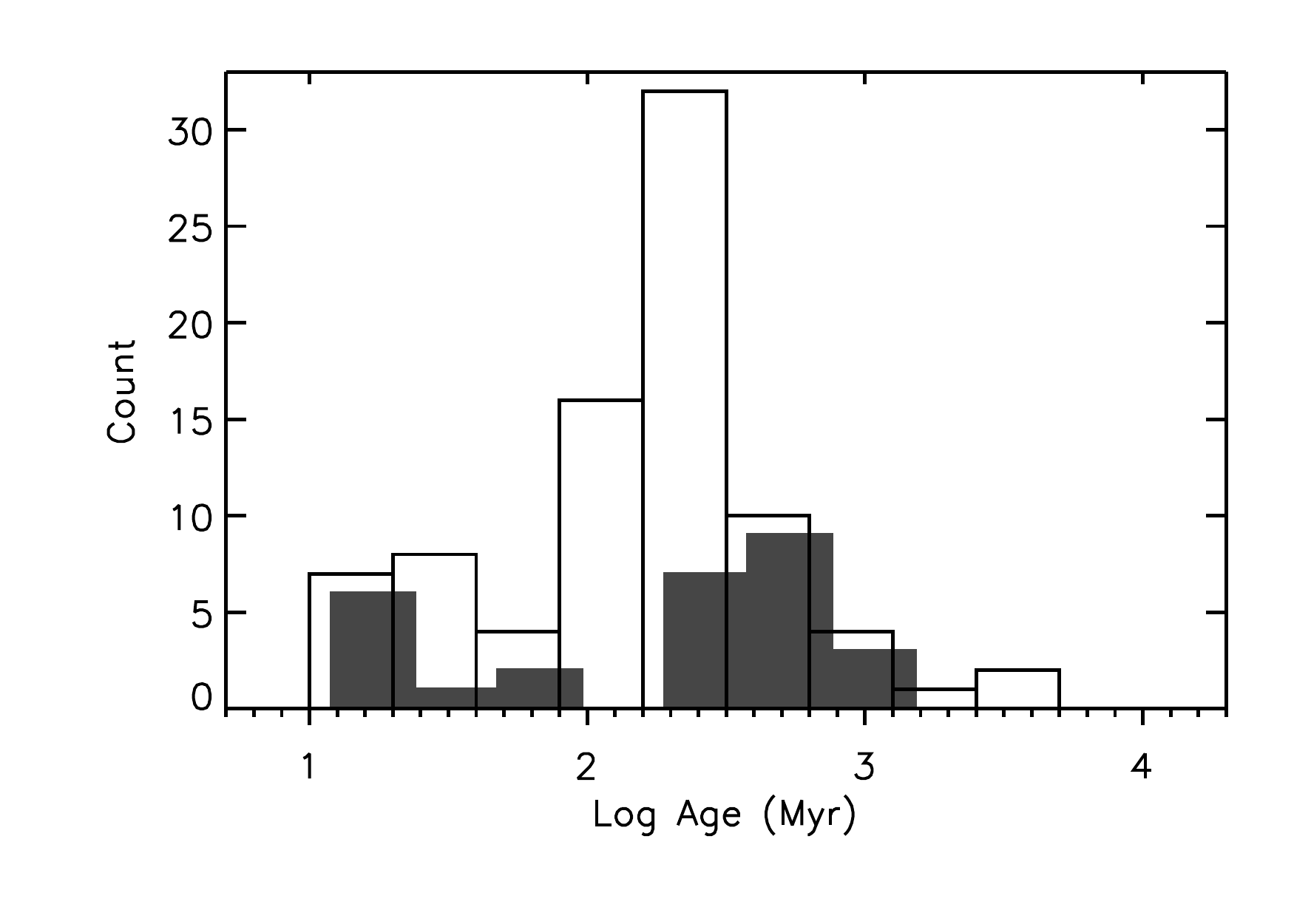}
\includegraphics[width=10cm,angle=0]{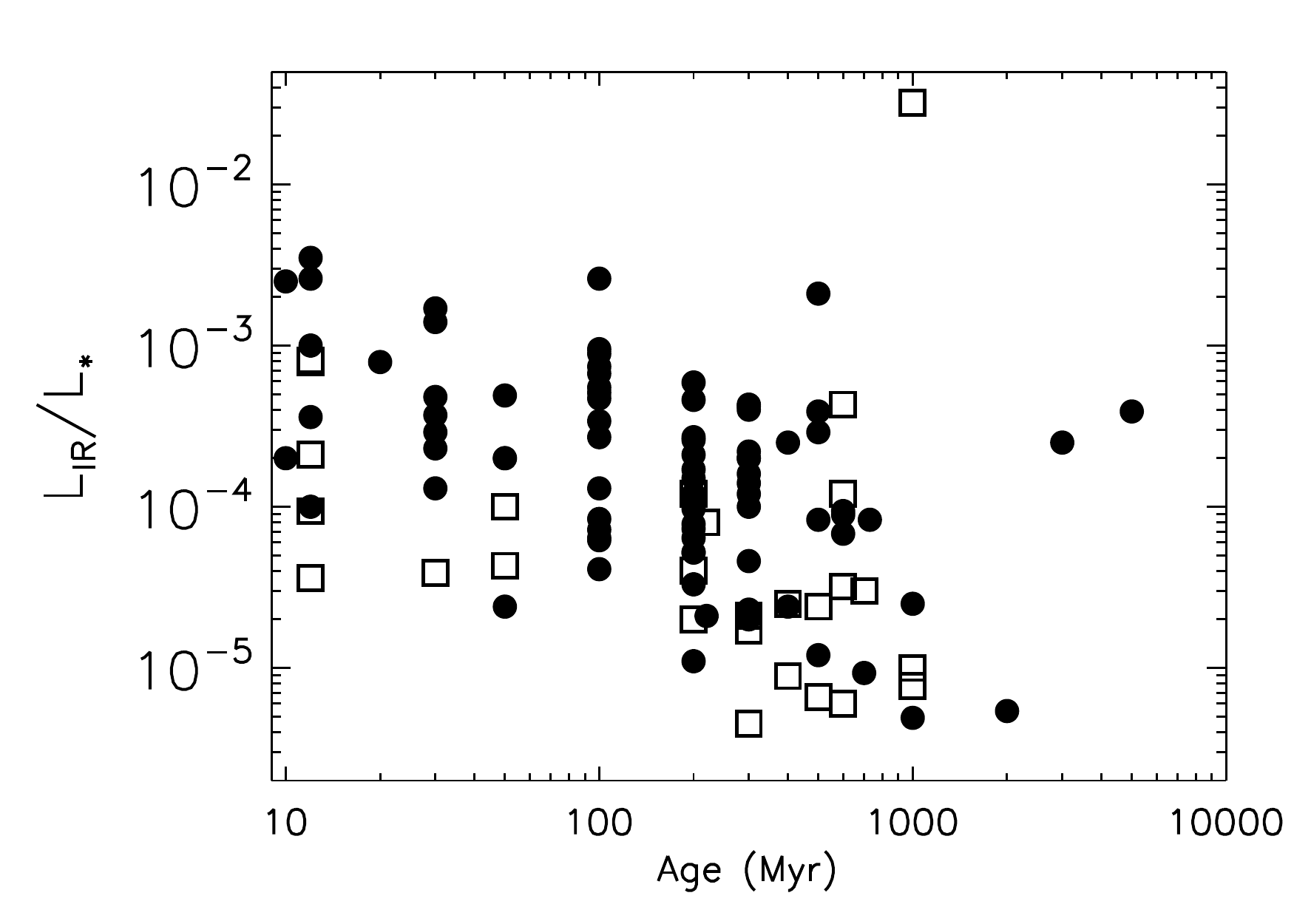}
\end{center}
\caption{
Age distribution in our debris disk sample. Open bars represent single systems, while the shaded bars represent multiples. While the histogram (top) suggests more multiple systems are older compared to the single stars, the plot of fractional luminosity vs.\ age (bottom; circles are single stars, squares are multiple systems) shows that at any given age, multiple stars are more likely to have lower fractional luminosities compared to single stars.
The $\sim$1~Gyr-old binary system with $L_\text{IR}/L_*\sim0.03$ is BD~+20~307.
}
\label{fig:debrisages}
\end{figure}

The difference in fractional luminosity may imply that the dust is being cleared more readily in multiple systems.
However, it may also be that there is a difference in the age of single and multiple systems in our sample.
Older systems are known to have disks with lower fractional luminosities \citep{Zuckerman:2001,Rhee:2007,Wyatt:2008}.
We show the distribution of ages in Figure~\ref{fig:debrisages}.
At a glance, the figure suggests our multiples are older, which is consistent with the results in \citet{Trilling:2007} in that the incidence of debris disks around multiples is marginally higher than that for single stars at ages $>$600~Myr.
However, closer examination (the bottom panel of Figure~\ref{fig:debrisages}) reveals that at any given age, single stars have higher fractional luminosities than multiple stars.
Hence, age alone cannot account for the lower fractional luminosities observed in multiple systems.
If the incidence of dust around multiple stars relative to its incidence among single stars is in fact more likely for older stars, as suggested by our results and those of \citet{Trilling:2007}, then an explanation may reside in the long-term orbital stability of rocky objects. 
Studies have demonstrated that planetesimal orbits can be disrupted by the gravitational influence of planets in the system as they migrate (see, for example, \citealt{Gomes:2005}; and references therein). 
However, as in our solar system, planetary systems should relax as they age and become progressively more dynamically quiescent. In contrast, the destabilizing influence of the gravity of stellar companions never really goes away.
The very dusty system BD+20~307, where rocky planets orbits were likely altered following a Gyr of evolution, probably contains three stellar members \citep{Zuckerman:2008b}.

\section{Separations} \label{debris:separations}

While the period distribution is suggestive, it is instructive to compare the stellar separation with that of the dust's orbital semi-major axis.
As previously mentioned, all the source references for our sample calculate dust semi-major axes from fits to the spectral energy distribution (SED) assuming blackbody dust grains.
Figure~\ref{fig:debrisdust} plots the stellar separations and dust semi-major axes for Table~\ref{tab:multiples} stars and those from \citet{Trilling:2007}.
Only the shortest separation for multiples is plotted with the exception of HIP~81641, where the dust is located around the single primary.
The grey region denotes dust locations that would be unstable based on the binary separation.
This is adopted from \citet{Trilling:2007} and is based on the stability study performed by \citet{Holman:1999}.
The stability of a test particle is defined by a critical semi-major axis, which is the maximum or minimum distance (depending on whether it's a circumstellar or circumbinary orbit) at which a test particle survives for $10^4$ times the binary period \citep{Holman:1999}.
The stability depends on the binary mass ratio and eccentricity, but we adopt average values of 0.5 for both and use Equations 2 and 5 in \citet{Holman:1999} to find that the critical semi-major axis is 0.12 and 3.8 times that of the binary separation.
This is comparable to the values of 0.15 and 3.5 used in \citet{Trilling:2007}.
We note that for circular orbits, the range in critical semi-major axis is narrower: 0.27 and 2.3 times that of the binary separation, and as such we adopt the more conservative estimate with $e=0.5$ in Figure~\ref{fig:debrisdust}.
In this regime, the companion would quickly disrupt the orbits of particles be they dust or larger objects.
We note that the actual boundaries of this unstable region are fuzzy for several reasons: (1) as already noted they depend on individual stellar parameters, (2) only gravitational forces are considered (ie, no radiative effects), (3) mean-motion resonances can disrupt test particles within an ostensibly stable region, and (4) several of our systems are known triples, whereas the \citet{Holman:1999} study applies for binaries.

Among our sample, there are 5 binaries and 1 triple that lie in the unstable region.
In addition, 3 systems from \citet{Trilling:2007} lie in that region (HD~46273, HD~80671, HD~127726), as described in their paper. Those three systems (two are triples, one is a quintuple system) have detected excesses only at 70\micron\, (in one case only marginal).
The ratio of dust to star separation for these 9 systems is between about 0.3 and 2.6.
These are listed in Table~\ref{tab:unstable}, where the separation listed for these systems is the conflicting one. 
As \citet{Trilling:2007} suggest, these systems may be undergoing a transient event, where dust has being generated by larger objects located farther out and has now migrated inward via Poynting-Robertson drag to its observed location.
In addition, both the radiation and gravity fields for triple or higher order multiples may be sufficiently complex to affect the dust temperature or orbital configuration of asteroids or planetesimals. 
%However, we describe below a number of effects that can potentially shift systems out of the unstable region in Figure~\ref{fig:debrisdust}.
However, we describe below a number of effects that can potentially make systems that actually are stable instead appear to be in the unstable region in Figure~\ref{fig:debrisdust} when viewed by a distant observer.

We note that with the exception of HD~1051, the systems listed in Table~\ref{tab:unstable} have infrared excesses detected at only 60 or 70\micron\, with detections at shorter wavelengths, such as 24 and 25\micron, being consistent with the stellar photosphere. Cooler dust temperatures and thus larger semi-major axis are consistent with the data. For clarity in Figure~\ref{fig:debrisdust}, only systems with a single detected excess at 70\micron\, have right pointing arrows.
Ill-determined dust temperature and locations are likely to be the single most important reason for systems to lie in the unstable zone. That is, the systems may actually not be unstable: the dust may be farther away and cooler than anticipated. This would shift systems towards the right in Figure~\ref{fig:debrisdust}.
Additional observations at longer wavelengths, such as with the Stratospheric Observatory for Infrared Astronomy (SOFIA) and the Herschel Space Observatory, will be key in determining temperatures for these systems.

An additional potential source of error in the horizontal placement of systems in Figure~\ref{fig:debrisdust} is the assumption of blackbody grains.  Small grains radiate inefficiently and, at a given distance from a star, will have higher temperatures than blackbody grains (\citealt{Zuckerman:2001}; and references therein).  As such, at a given temperature small grains will be located further to the right on the Figure~\ref{fig:debrisdust} horizontal axis than will blackbody grains.  A handful of systems have been resolved in thermal emission (where the dust particles themselves radiate at wavelengths between 20 and 1000~$\micron$), or in scattered light, where small grains reflect and scatter light from the central star.  We consider these systems in Table~\ref{tab:resolved} and Figure~\ref{fig:resolved}, noting that a handful of known resolved systems are not listed. These include $\beta$~Pic, $\tau$~Ceti, $\gamma$~Oph, and $\zeta^2$~Ret, as either too little information is available or, in $\beta$~Pic's case, too much, and thus meaningful determination of a characteristic resolved radius is not possible.  Note that HD~141569A and HD~98800B appear in Table~\ref{tab:resolved}; these usually have been classified as transition disks.  For Table~\ref{tab:resolved} and Figure~\ref{fig:resolved} we use an average for the range in location of the dust or adopt the location of peak emission.  Because grain semi-major axes plotted in Figure~\ref{fig:debrisdust} are based on thermal emission, the relevant comparison of semi-major axes in Table~\ref{tab:resolved} and Figure~\ref{fig:resolved} is with spatially resolved thermal emission and not with scattered light.  Thermally resolved disk radii range from one to five times larger than expected from blackbody-fit radii and likely are due to the particular composition, size, and porosity of grains in the disk.  For actual disk semi-major axes 5 times larger than the blackbody model, one can shift at most 5 systems out of 9 from the unstable region in Figure~\ref{fig:debrisdust}.  Hence consideration of the actual sizes of disks would result in shifting a few systems to the right (larger disk radii) and into stability.  Similar considerations could shift a few apparently stable systems into an unstable configuration.

Because of projection effects and location along the orbit, the observed binary separation is a lower limit and could be somewhat larger than that plotted in Figure~\ref{fig:debrisdust}. 
The ratio between projected separation ($\rho$) and semi-major axis ($a$) for a circular orbit is:
\begin{align*}
\rho = a \sqrt{1 - \sin^2 \omega \sin^2 i}
\end{align*}
where $\omega$ is the location along the orbit and thus has a total range of 0 to $2\pi$ (from \citealt{Macintosh:1994}; however, see also \citealt{Leinert:1993}).
The inclination, $i$, has a range of 0 to $\pi/2$ ($i=0$ denotes a face-on orbit).
We can estimate the average ratio:
\begin{align*}
<\frac{\rho}{a}> = \frac{\displaystyle\int^{\pi/2}_0 \int^{2\pi}_0 \sqrt{1 - \sin^2 \omega \sin^2 i}\, di d\omega }{\displaystyle\int^{\pi/2}_0 \int^{2\pi}_0 \, di d\omega } \approx 0.842 \approx \frac{1}{1.19}
\end{align*}
So the average binary separation could be $\sim20$\% larger than what is actually measured.
However, on an individual-case basis, objects may have inclinations and orbit locations such that objects move outside the unstable region. We generated 100,000 random inclinations between 0 and 90$^\circ$ and orbit locations between 0 and 360$^\circ$ to compute the ratio of projected to actual separation. This distribution is integrated and normalized in order to estimate how often ratios lower than a particular value appear. 
For the unstable systems listed in Table~\ref{tab:unstable}, we estimate what semi-major axis is required to yield a dust radius-to-star separation ratio of 0.12 and calculate the probability (the final column) of achieving stability in Figure~\ref{fig:debrisdust} based on the measured stellar separation.
%This probability for our unstable systems is quoted in the last column of Table~\ref{tab:unstable}.
In all cases the likelihood is low that the semi-major axis of the binary system is much larger than the observed binary separation.  Thus, at most one unstable disk system would be shifted upward in Figure~\ref{fig:debrisdust} into a stable configuration.   However, as noted in the first paragraph of the present Section, Figure~\ref{fig:debrisdust} assumes a binary eccentricity of 0.5.  If the HD~1051 binary were instead to have a circular orbit, then the probability that the orbiting dust is stable would be 60\%.  In addition, should the HIP~76127 binary also possess a circular orbit, then the circumbinary dust would be stable as its dust radius-to-star separation ratio is larger than  the critical semi-major axis for circular orbits (2.3).
In practice, as most binary orbits are eccentric, the companion will spend more time away from the primary, resulting in a lower ratio between projected separation and semi-major axis than the average estimated here.

Summarizing the above considerations, data points in Figure~\ref{fig:debrisdust} may be shifted  as a result of incorrect temperature determination, dust grain characteristics, and projection effects. The dominant effect in our sample is likely the incorrect determination of dust temperatures, and thus, of dust semi-major axes.
%projection effects will shift some of the plotted data points in Figure~\ref{fig:debrisdust} upward, while considerations of resolved disk sizes would shift some points towards the right; the exact amount, however, will depend on a per-object basis.
The other two effects are, in general, probably insufficient to shift all systems outside the region where dust particle orbits are expected to be unstable.
The result of binary interactions in the disks will be to either to clear gaps or to truncate the disks \citep{Lubow:2000,Artymowicz:1994}.
The small number of these disk systems in the unstable zone suggests that binaries, although they can form disks, are more likely to disrupt the formation and evolution of planetary systems when their separations are comparable to typical disk sizes.

\section{Disk Masses}\label{diskmass}

A final comparison one can make for our debris disk sample are dust masses determined from (sub)millimeter photometry.
\citet{Nilsson:2010} have summarized all submm and longer wavelength detections of debris disks, but only briefly comment on binarity in their paper.
We have compared our sample to their list, but only 23 of our debris disks have submm or mm measurements.
Of those 23 only 3 are binaries and thus our comparisons are limited by small number statistics.
Disk dust masses can be readily estimated with:
\begin{equation*}
M_d = \frac{S_\nu D^2}{k_\nu B_\nu(T_d)}
\end{equation*}
where $S_\nu$ is the submm flux, $D$ is the distance between the dust and Earth, and $B_\nu$ is the blackbody flux at the dust temperature $T_d$.
The dust opacity, $k_\nu$, is extrapolated from an opacity of 1.7 cm$^2$ g$^{-1}$ at 880~$\mu$m assuming a dust opacity slope $\beta$=1: $k_\nu = 1.7 (880/\lambda)^\beta$ (see \citealt{Zuckerman:2008a}).

We list our estimated disk dust masses in Table~\ref{tab:masses}.
The average dust mass for our 20 single stars is $\sim7$ Moon masses ($M_\text{Moon}$; the median is $\sim$3~$M_\text{Moon}$).
In contrast, our 3 binary star systems have an average disk mass of $\sim0.7$~$M_\text{Moon}$.
%: HIP~27072 (0.04~$M_\text{Moon}$), HIP~107649 (0.5), and Fomalhaut (1.7).
These single and binary star dust masses, while few, are consistent with our results for fractional luminosities and dust separations.
That is, binaries or multiples are more effective at clearing their disks and as such, they possess lower disk masses (as previously found for very young systems; see \citealt{Jensen:1996}).
We note, however, that these few binaries have ages $\geq200$~Myr, whereas our single stars are predominantly younger than $\sim200$~Myr.
Hence, this disk mass difference, while suggestive, must be taken with care.

Further submm and mm observations of our debris disk sample, for example with JCMT, APEX, or ALMA, will be key in gathering a larger sample of disk mass measurements for both single and multiple debris disk systems.
In particular, for the systems in the unstable region of Figure~\ref{fig:debrisdust}, we extrapolate what flux density is expected at 850\micron.
We use the systems measured by \citet{Williams:2006} and compare the $F_{850}/F_{60}$ ratio as a function of dust temperature in order to estimate $F_{850}$.
We find that our unstable disks listed in Table~\ref{tab:unstable} (not including the three \citealt{Trilling:2007} systems), are anticipated to have flux densities of $\sim1-10$~mJy at 850\micron.
In addition, these systems should have angular radii (from the SED fits) of $\sim0.4-4$\arcsec.
In a single hour, at 345~GHz with a 1\arcsec\, beam, full ALMA (54 antennae) can reach sensitivities of $\sim\!0.02$~mJy/beam, making systems like these readily detectable and resolvable.

\section{Conclusions} \label{conclusions}

We find that the fraction of stars in binary or multiple systems among our debris disk systems is $25\pm4\%$. This is less than the anticipated $\sim\!1/2$ of multiple stars and could be due either to a physical difference (ie, less multiples among debris disk systems) or to incomplete multiplicity data for some of our stars. 
We performed an adaptive optics search on 20 not previously well-studied systems in order to search for binaries in the 10--1000~AU separation range.
We discovered only a single binary: HD~1051, an A-star with an M-star companion with projected separation of 628~AU.
The distribution of systems in stellar separation vs.\ blackbody dust semi-major axis confirms the theoretical expectation that dust will not form or survive at separations comparable to that of the semi-major axis of the secondary star.
Additionally, the fractional luminosities and disk masses, the latter inferred from a handful of submm detections, are lower for the multiples systems; this serves to strengthen the idea that these systems clear out their disks more effectively than do single stars.

What does this imply for planet formation?
First, many binaries do possess disks and as such could very well form planets.
However, our results suggest that binaries are less likely to possess long-lived, massive disks.
We infer this by our result of fewer than expected binaries in our sample and the lower disk masses and fractional luminosities among the binaries.
%In addition to this, our results show that binaries are unlikely to possess disks whose radii are comparable to the orbital separations of the two or more stars.
The fact that the binary semi-major axis distribution peaks around $\sim$30~AU \citep{Duquennoy:1991} suggests that most binaries will only rarely form planets at semi-major axes comparable to the gas giants in our own solar system.
For a binary to possess planets it must either be a very widely separated binary with planets orbiting a single star (in fact, $\sim$20\% of known extrasolar planets are in such systems; see \citealt{Raghavan:2006}) or it must be a very closely separated binary with planets forming in a  massive circumbinary disk.

\begin{figure}[htb]
\begin{center}
\includegraphics[width=12cm,angle=0]{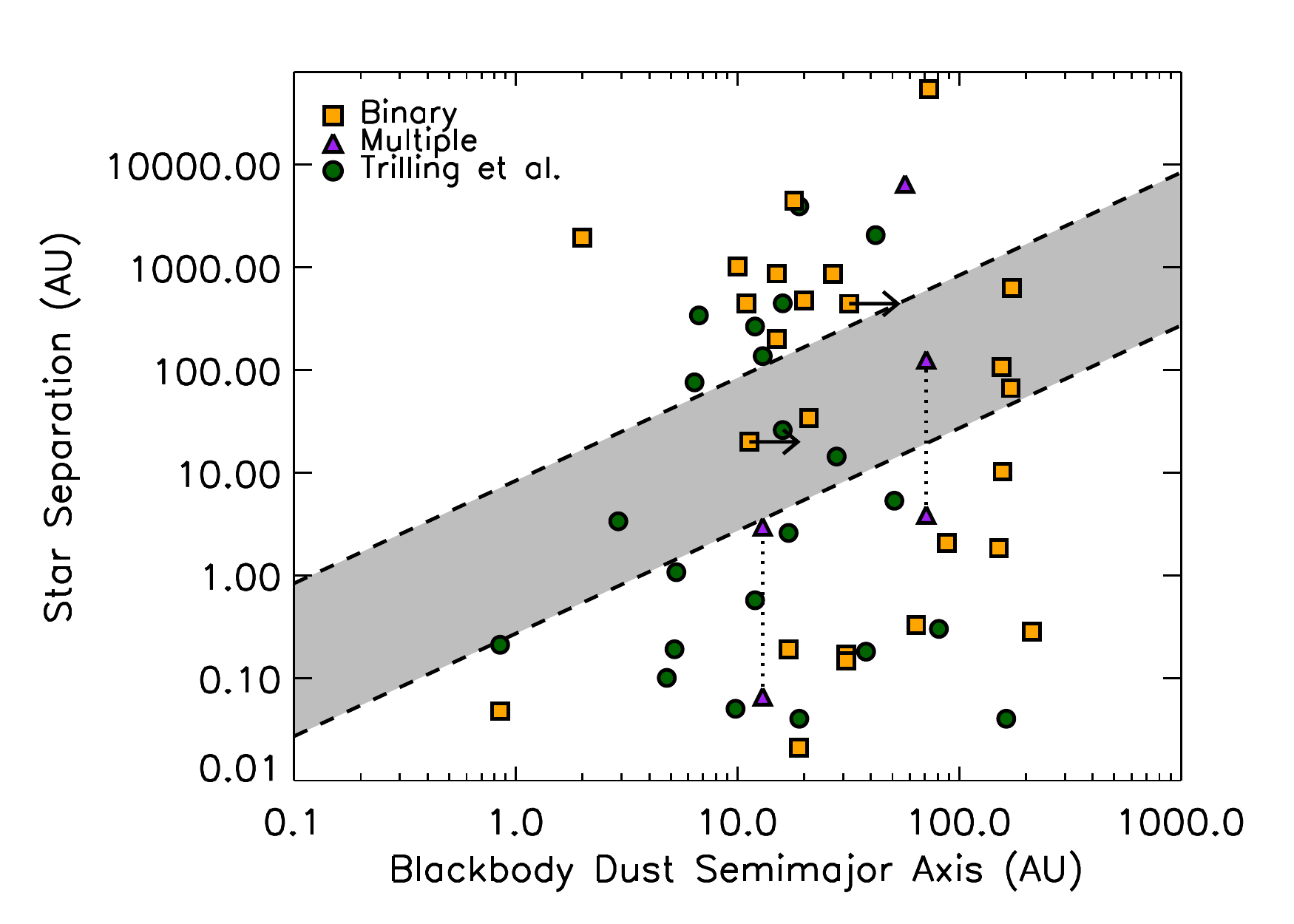}
\end{center}
\caption{Stellar separation versus blackbody dust semi-major axis for debris disk stars including those from \citet{Trilling:2007}.
%Only the shortest period is used for systems with multiplicities higher than 2 (with the exception of HIP~81641 where the disk may be around the single primary so the separation of the companion close binary is not relevant).
Triple systems are plotted at both orbital separations with a dotted line connecting them; the exception is HIP~81641 where the disk may be around the single primary so the separation of the companion close binary is not relevant.
The two systems with right pointing arrows are HIP~66704 and HIP~92680 (PZ~Tel), for which dust emission is detected only at $70\micron$; for clarity we do not label the systems with excesses detected only at 60\micron\, (see Section~\ref{dust_T}).
The grey region in the plot highlights where the gravitational field of a companion is expected to significantly affect dust in the system (see the first paragraph of Section~\ref{debris:separations})}
\label{fig:debrisdust}
\end{figure}

\begin{figure}[htb]
\begin{center}
\includegraphics[width=12cm,angle=0]{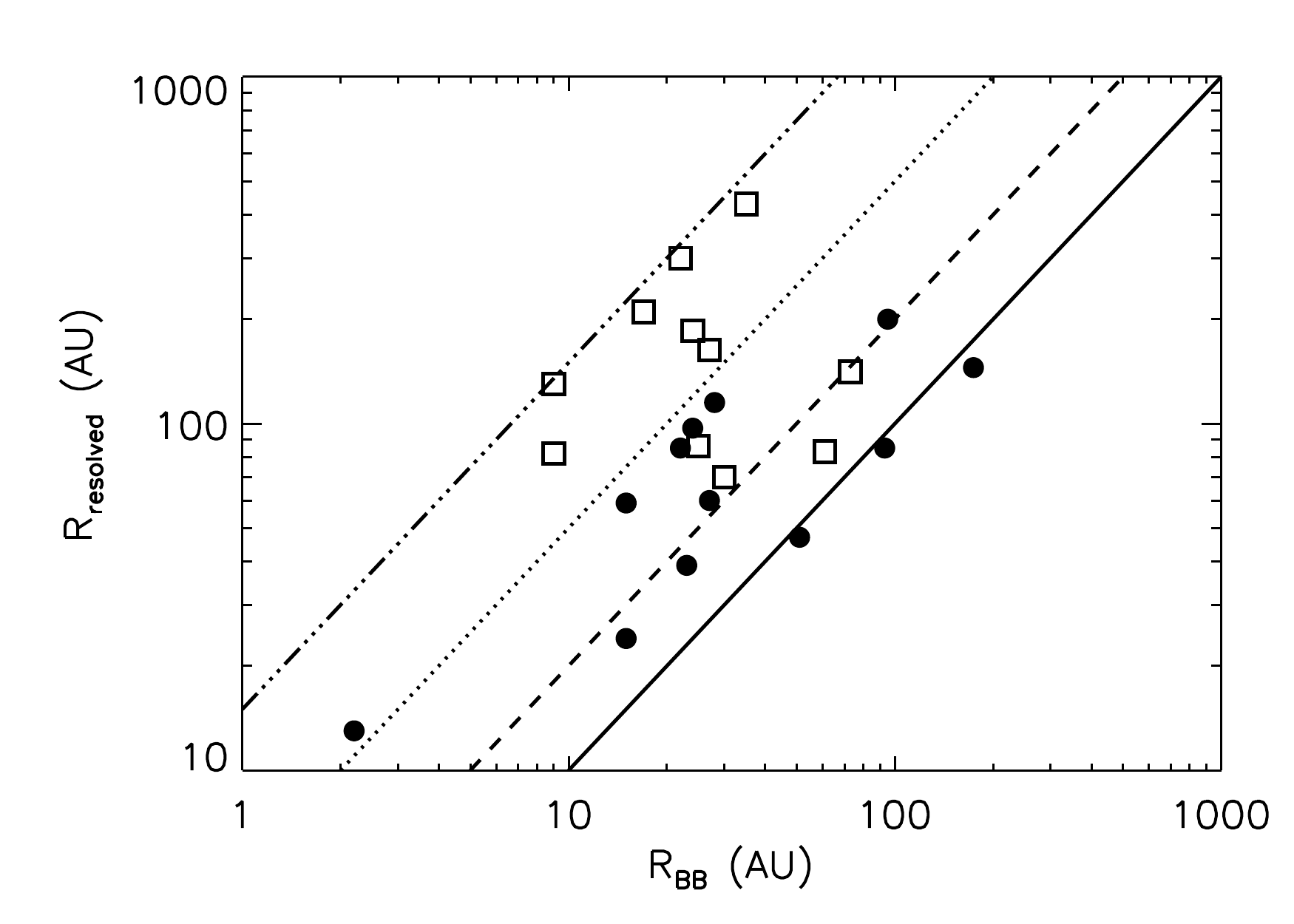}
\end{center}
\caption{
Resolved vs black body disk radii for debris disks as listed in Table~\ref{tab:resolved}.
The lines represent ratios of 1:1 (solid), 2:1 (dashed), 5:1 (dotted), and 15:1 (triple-dot-dashed) when comparing resolved disk sizes to those inferred from blackbody fits. 
Circles are disks resolved in thermal emission, while squares are disks resolved in scattered light. 
Thermally resolved disks have sizes $\sim1-5$ times that estimated from the blackbody SED, whereas disks resolved in scattered light tend to be more extended.
}
\label{fig:resolved}
%\epsscale{0.4}
\end{figure}

Results from unbiased Herschel surveys followed by ALMA imaging should address uncertainties associated with Figures~\ref{fig:debrisdust} and \ref{fig:resolved} of the present study.

\acknowledgements
We thank Gaspard Duchene and Harold Butner for useful discussions and for allowing us to observe some Table~\ref{tab:debris_sample} systems in our joint observing runs.
Many thanks to Sasha Hinkley and Lewis Roberts for double checking our targets with their binary program, to Marshall Perrin for his IDL IRCAL routines, and to Joseph Rhee for assistance with determining dust temperatures.
We thank our referee, Michal Simon, for a prompt review and constructive suggestions.
This research has made use of the Washington Double Star Catalog maintained at the U.S. Naval Observatory, and the SIMBAD database and the VizieR catalog access tool, operated at CDS, Strasbourg, France.
This research was supported in part by NASA grants to UCLA.

\bibliographystyle{apj}
%\bibliography{papers}{}

\clearpage

\input{debris}
\clearpage
\input{debris_binaries_singlespace}
\clearpage

\begin{deluxetable}{cclcc}
\tablecaption{Multiplicity Fractions\label{debris:binfraction}}
\tablewidth{0pt}
\tablehead{
\colhead{Spectral Type} & \colhead{Number} & \colhead{Percent}
& \colhead{From ET08} & \colhead{From DM91}
}
\startdata
B & $5/6$ & $83^{+6}_{-23}$ & \nodata & \nodata \\
A & $12/47$ & $26^{+7}_{-5}$ & 46.0 & \nodata \\ 
F & $5/41$ & $12^{+7}_{-3}$ & 47.4 & \nodata \\
G & $4/12$ & $33^{+15}_{-10}$ & 45.0 & 57\\
K & $2/5$ & $40^{+21}_{-16}$ & 29.1 & \nodata \\
\enddata
\tablecomments{Fraction of multiple stars (all as percentages) broken down by spectral type of the primary star with comparisons from the literature (ET08: \citealt{Eggleton:2008}; DM91: \citealt{Duquennoy:1991}). %Errors are binomial errors estimated as described by \cite{Burgasser:2003}.
%Not listed are the B-type stars from Table~\ref{tab:debris_sample} of which 5/6 ($83^{+6}_{-23}$\%) are multiples.
}
\end{deluxetable}

% Lick status table
\begin{table}[ht!]
\caption{IRCAL Debris Disk Observations}
\label{tab:ircalobs}
\begin{center} {\footnotesize
\begin{tabular}{llll}
\hline
\hline
 Name & UT Date & Filter & Exposure Time (s)\\
\hline
HIP~63076 & 2009 June 7 & Ks & 11.4\\
HIP~70952 & 2009 June 7 & J,H,Ks & 20, 11.4, 11.4\\
HIP~71284 & 2009 June 7 & BrG-2.16 & 20\\ %F039
HIP~76635 & 2009 June 7 & Ks & 20\\
HIP~78554 & 2009 June 7 & Ks & 11.4\\
HIP~81800 & 2009 June 7 & Ks & 11.4\\
HIP~83480 & 2009 June 7 & Ks & 20\\
HIP~85157 & 2009 June 7 & Ks & 11.4\\
HIP~85537 & 2009 June 7 & Ks & 11.4\\
HIP~87108 & 2009 June 7 & BrG-2.16 & 40\\
HIP~106741 & 2009 June 7 & Ks & 40\\
HIP~5626 & 2009 October 29 & Ks & 52\\
HIP~15197 & 2009 October 29 & BrG-2.16 & 52\\ %A064
HIP~10670 & 2009 October 29 & BrG-2.16 & 26\\ %A067
HIP~32480 & 2009 October 29 & BrG-2.16 & 56\\ %F044
HIP~35550 & 2009 October 29 & BrG-2.16 & 54\\ %F067 %known binary
HIP~41152 & 2009 October 29 & Ks & 26\\
HIP~41307 & 2009 October 29 & BrG-2.16 & 52\\ %A082
HIP~69732 & 2010 August 3 & J,Fe~II,BrG-2.16 & 6, 4.2, 3 \\ %A053
HIP~1185 & 2010 August 4 & Ks & 20\\
\hline
\end{tabular} }
\end{center}
%{BrG-2.16 and Fe~II are narrow band filters centered on 2.167 and 1.644\micron, respectively. They are similar to the Ks and H filters, but $\sim10$ times narrower.}
\end{table}

\begin{deluxetable}{ccrrrrrr}
\tablecaption{IRCAL Calibrations\label{tab:wdscalib}}
\tablewidth{0pt}
\tablehead{
\colhead{Name} & \colhead{Grade} & \colhead{PA}
& \colhead{$\rho$} & \colhead{$\Delta$X}
& \colhead{$\Delta$Y} & \colhead{Calc. PA} & \colhead{Calc. $\rho$} \\
\colhead{} & \colhead{} & \colhead{(deg)}
& \colhead{(\arcsec)} & \colhead{(pixels)}
& \colhead{(pixels)} & \colhead{(deg)} & \colhead{(\arcsec)}
}
\startdata
HD 212698 & 4	  & 41.4 & 1.30 & -11.76 & 11.98 & 43.1 & 1.26\\ %G107
HD 165341 & 1 & 130.8 &	5.77	& -60.70 & -47.76 &	130.6 & 5.76	\\
GJ 65 & 3   & 43.1 & 1.96 & -18.72 & 19.98 & 41.8 & 2.06 \\ %M004
HD 38 & 4  & 186.6 & 5.94 & 7.73 & -77.12 & 185.9 & 6.00 \\
HD 38 & 4  & 186.2 & 6.04 & 7.73 & -77.12 & 185.9 & 6.00 \\
HD 133640 & 2  & 60.1 & 1.54 & -18.40 & 9.89 & 60.7 & 1.54 \\ %G025
HD 160269 & 3  & 315.7 & 0.99 & 9.32 & 8.73 & 315.5 & 0.96 \\ %G034
HD 146361 & 4  & 237.5 & 7.12 & 81.90 & -49.00 & 237.9 & 7.06 \\ %G122
\enddata
\tablecomments{Stars used as part of our IRCAL calibration, as described in Section~\ref{hip1185}. Grade denotes the quality of the orbit, with lower numbers representing better orbits. 
The two columns after Grade denote the position angle (PA) and separation ($\rho$) for the system at the time of our observations. %(from the Sixth Catalog of Orbits of Visual Binary Stars); two orbits were provided for HD~38.
$\Delta X$ and $\Delta Y$ are the pixel separation in the X and Y direction and the two columns following that are the calculated position angle and separation using the best-fit solution in Section~\ref{hip1185}. 
}
\end{deluxetable}

% Unstable Systems
\begin{table}[h]%[htdp]
\caption{Unstable debris disk systems}
\begin{center} {%\scriptsize
\begin{tabular}{lccccc}
\hline
\hline
Name & Sep. & $R_\text{dust}$ & Ratio & Type & Stability \\
 & (AU) & (AU) & & & (\%) \\
 \hline
HD~1051		& 628	& 173	& 0.28	& binary 	& 6 \\ 
HIP 35550	& 126	& 71		& 0.57	& triple 	& 1 \\
HIP 42430	& 34		& 21		& 0.62	& binary 	& 1 \\
HIP 76127	& 66		& 171	& 2.58	& binary 	& 0.1 \\
HIP 90185	& 106	& 155	& 1.46	& binary 	& 0.2 \\
HIP 92680	&  20	& 11		& 0.57	& binary 	& 1 \\
\hline
HD~46273	&  26	&  16	& 0.62	& quintuple & 1 \\
HD~80671	&  3.4	&  2.9	& 0.87	& triple	& 0.6 \\
HD~127726	&  14	&  28	& 1.96	& triple	& 0.1\\
\hline
\end{tabular} }
\end{center}
{\small List of unstable systems with the ratio of the dust to star separation. For completeness, we include the 3 systems from \citet{Trilling:2007} at the bottom of the list. The final column denotes the probability that the inclination and location of the secondary star along a circular orbit are such that in actuality the system lies outside of the unstable zone (see Section~\ref{debris:separations} for details). 
%HIP~92680 and the systems from \citet{Trilling:2007} have detected infrared excess only at 70\micron. HIP~35550, HIP~42430, HIP~76127, and HIP~90185 have detected excess only at 60\micron. As such, the dust in these systems may well have larger semi-major axes than indicated in the table.
}
\label{tab:unstable}
\end{table}

\clearpage
\input{resolved_disks}

\clearpage
% Dust Masses
\begin{table}[h]%[htdp]
\caption{Debris disk dust masses}
\begin{center} {%\scriptsize
\begin{tabular}{lccccccc}
\hline
\hline
Name  & $\lambda_\text{dust}$ & $F_\nu$  & Mass  & Age\\
        & (\micron) & (mJy) & ($M_{moon}$) & (Myr) \\
\hline
AU Mic	&	850	& $	14.4	\pm	1.8	$ &	0.5	&	12	\\
$\beta$ Pic	&	1300	& $	24.9	\pm	2.6	$ &	5.8	&	12	\\
HIP 95270	&	870	& $	51.7	\pm	6.2	$ &	36.5	&	12	\\
HIP 99273	&	350	& $	54	\pm	15	$ &	2.2	&	30	\\
HIP 114189	&	850	& $	10.3	\pm	1.8	$ &	6.3	&	30	\\
HIP 16449	&	870	& $	17.6	\pm	8	$ &	33.1	&	50	\\
HIP 10670	&	850	& $	5.5	\pm	1.8	$ &	1.8	&	100	\\
HIP 36948	&	350	& $	95	\pm	12	$ &	2.5	&	100	\\
HIP 11360	&	850	& $	4.9	\pm	1.6	$ &	2.9	&	100	\\
HIP 60074	&	850	& $	20	\pm	4	$ &	5.7	&	100	\\
HIP 22226	&	870	& $	6.9	\pm	5	$ &	13.4	&	100	\\
HIP 87108	&	870	& $	12.8	\pm	5.2	$ &	2.6	&	200	\\
HIP 101612	&	870	& $	13	\pm	7.1	$ &	3.2	&	200	\\
HIP 6878	&	1200	& $	3.2	\pm	0.9	$ &	4.5	&	200	\\
HIP 90936	&	870	& $	18	\pm	5.4	$ &	9.7	&	200	\\
Vega	&	1300	& $	11.4	\pm	1.7	$ &	0.6	&	220	\\
$\eta$ Crv	&	850	& $	7.5	\pm	1.2	$ &	0.3	&	300	\\
HIP 32480	&	850	& $	5.5	\pm	1.1	$ &	0.5	&	600	\\
$\epsilon$ Eri	&	1300	& $	24.2	\pm	3.4	$ &	0.4	&	730	\\
$\sigma$ Boo	&	850	& $	6.2	\pm	1.7	$ &	0.7	&	1000	\\
\hline			
Fomalhaut	&	850	& $	97	\pm	5	$ &	1.7	&	220	\\
HIP 107649	&	870	& $	5.1	\pm	2.7	$ &	0.5	&	600	\\
HIP 27072	&	850	& $	2.4	\pm	1	$ &	0.04	&	1000	\\\hline
\end{tabular} }
\end{center}
{\small Disk dust masses for Table~\ref{tab:debris_sample} stars listed in \citet{Nilsson:2010}. The last three systems are binaries. $\lambda_\text{dust}$ is the wavelength at which the dust is detected and the one used to estimate dust mass (see Section~\ref{diskmass}); dust temperatures and distances from Earth are listed Table~\ref{tab:debris_sample}.
}
\label{tab:masses}
\end{table}

\clearpage

\end{document}

%% file: debris.tex
\begin{deluxetable}{lllcccccccr}
\tablecolumns{11}
\tabletypesize{\scriptsize}
\rotate
\tablewidth{0pc}
\tablecaption{Debris Disk Sample}

\tablehead{
\colhead{Name} & \colhead{HIP} & \colhead{Spectral} & \colhead{Dist.} & 
\colhead{Detection} & \colhead{$T_\text{dust}$} & \colhead{$L_{IR}/L_*$} & \colhead{$R_\text{dust}$} & \colhead{Multiple?} & \colhead{Age} & \colhead{References}    \\
\colhead{} & \colhead{Number} & \colhead{Type} & \colhead{(pc)} & \colhead{} & \colhead{(K)} & 
\colhead{} & \colhead{(AU)} & \colhead{} & \colhead{(Myr)} & \colhead{} }

\startdata
HR 9	&	HIP 560	&	F2	&	39.1	&	MIPS	&	120	&	1.0E-04	&	10	&	N	&	12	&	\citealt{Rebull:2008}	\\
HD 432	&	HIP 746	&	F2	&	16.7	&	IRAS	&	120	&	2.5E-05	&	28	&	N	&	1000	&	\citealt{Rhee:2007}	\\
HD 1051	&	HIP 1185	&	A7	&	88.3	&	IRAS	&	40	&	4.3E-04	&	173	&	Y	&	600	&	\citealt{Rhee:2007}	\\
HD 2262	&	HIP 2072	&	A5	&	24	&	MIPS$^a$	&	140	&	1.1E-05	&	14	&	N	&	200	&	\citealt{Chen:2005}	\\
HD 6798	&	HIP 5626	&	A3	&	83.5	&	IRAS	&	75	&	1.5E-04	&	93	&	N	&	200	&	\citealt{Rhee:2007}	\\
HD 8538	&	HIP 6686	&	A5	&	30.5	&	IRAS &	85	&	6.0E-06	&	88	&	Y	&	600	&	\citealt{Rhee:2007}	\\
HD 8907	&	HIP 6878	&	F8	&	34.2	&	IRAS/MIPS/ISO	&	45	&	2.1E-04	&	59	&	N	&	200	&	\citealt{Rhee:2007}	\\
HD 9672	&	HIP 7345	&	A1	&	61.3	&	IRAS	&	80	&	7.9E-04	&	60	&	N	&	20	&	\citealt{Rhee:2007}	\\
HD 10472	&	HIP 7805	&	F2	&	66.6	&	IRAS/MIPS	&	70	&	3.7E-04	&	30	&	N	&	30	&	\citealt{Rhee:2007}	\\
HD 10647	&	HIP 7978	&	F8	&	17.4	&	IRAS	&	65	&	4.2E-04	&	22	&	N	&	300	&	\citealt{Rhee:2007}	\\
HD 10638	&	HIP 8122	&	A3	&	71.7	&	IRAS	&	85	&	4.7E-04	&	33	&	N	&	100	&	\citealt{Rhee:2007}	\\
HD 10939	&	HIP 8241	&	A1	&	57	&	IRAS/MIPS	&	75	&	6.4E-05	&	80	&	N	&	200	&	\citealt{Rhee:2007}	\\
%BD +20 307	&	HIP 8920	&	G0	&	91.9	&	IRAS/Gemini/Keck	&	435	&	4.0E-02	&	0.5	&	Y	&	1000	&	\citealt{Song:2005,Weinberger:2008}	\\
BD +20 307	&	HIP 8920	&	G0	&	91.9	&	IRAS/Gemini/Keck	&	358	&	3.2E-02	&	0.85	&	Y	&	1000	&	\citealt{Song:2005,Weinberger:2011}	\\
HD 14055	&	HIP 10670	&	A1	&	36.1	&	IRAS	&	75	&	7.2E-05	&	80	&	N	&	100	&	\citealt{Rhee:2007}	\\
HIP 10679	&	HIP 10679	&	G2	&	34	&	MIPS	&	100	&	8.0E-04	&	20	&	Y	&	12	&	\citealt{Rebull:2008}	\\
HD 15115	&	HIP 11360	&	F2	&	44.8	&	IRAS/MIPS/ISO	&	65	&	5.1E-04	&	35	&	N	&	100	&	\citealt{Rhee:2007}	\\
AG Tri	&	HIP 11437	&	K8	&	42.3	&	MIPS	&	65	&	7.9E-04	&	10	&	Y	&	12	&	\citealt{Rebull:2008}	\\
HD 15745	&	HIP 11847	&	F0	&	63.7	&	IRAS/MIPS/ISO	&	85	&	1.7E-03	&	22	&	N	&	30	&	\citealt{Rhee:2007}	\\
HD 16743	&	HIP 12361	&	F1	&	60	&	IRAS/MIPS	&	40	&	5.9E-04	&	119	&	N	&	200	&	\citealt{Rhee:2007}	\\
HD 17390	&	HIP 12964	&	F3	&	45.1	&	IRAS/MIPS	&	55	&	2.0E-04	&	55	&	N	&	300	&	\citealt{Rhee:2007}	\\
HD 17848	&	HIP 13141	&	A2	&	50.7	&	IRAS/MIPS	&	55	&	6.4E-05	&	97	&	N	&	100	&	\citealt{Rhee:2007}	\\
HD 19356	&	HIP 14576	&	B8	&	28.5	&	IRAS/MIPS	&	250	&	1.7E-05	&	13	&	Y	&	300	&	\citealt{Su:2006,Rhee:2007}	\\
HD 20320	&	HIP 15197	&	A5	&	36.8	&	IRAS/MIPS	&	95	&	2.5E-05	&	31	&	Y	&	400	&	\citealt{Rhee:2007}	\\
HD 21997	&	HIP 16449	&	A3	&	73.8	&	IRAS/MIPS	&	60	&	4.9E-04	&	82	&	N	&	50	&	\citealt{Rhee:2007}	\\
$\epsilon$ Eri	&	HIP 16537	&	K2	&	3.2	&	IRAS/MIPS/ISO	&	40	&	8.3E-05	&	27	&	N	&	730	&	\citealt{Rhee:2007}	\\
HD 25457	&	HIP 18859	&	F5	&	19.2	&	IRAS/MIPS/ISO	&	85	&	1.3E-04	&	16	&	N	&	30	&	\citealt{Rhee:2007}	\\
HD 27290	&	HIP 19893	&	F4	&	20.3	&	IRAS	&	80	&	2.3E-05	&	31	&	N	&	300	&	\citealt{Rhee:2007}	\\
HD 30447	&	HIP 22226	&	F3	&	78.1	&	IRAS/MIPS	&	65	&	8.9E-04	&	37	&	N	&	100	&	\citealt{Rhee:2007}	\\
HD 31295	&	HIP 22845	&	A0	&	37	&	IRAS/MIPS	&	80	&	8.4E-05	&	49	&	N	&	100	&	\citealt{Rhee:2007}	\\
HD 34324	&	HIP 24528	&	A3	&	85.8	&	IRAS	&	100	&	1.7E-04	&	28	&	N	&	200	&	\citealt{Rhee:2007}	\\
HR 1817	&	HIP 25486	&	F7	&	26.8	&	MIPS$^a$	&	80	&	3.6E-05	&	17	&	Y	&	12	&	\citealt{Rebull:2008}	\\
HD 37484	&	HIP 26453	&	F3	&	59.5	&	IRAS/MIPS	&	90	&	2.9E-04	&	19	&	N	&	30	&	\citealt{Rhee:2007}	\\
HD 38206	&	HIP 26966	&	A0	&	69.2	&	IRAS/MIPS	&	85	&	2.0E-04	&	53	&	N	&	50	&	\citealt{Rhee:2007}	\\
HD 38393	&	HIP 27072	&	F7	&	9	&	IRAS/MIPS	&	90	&	7.7E-06	&	15	&	Y	&	1000	&	\citealt{Rhee:2007}	\\
HD 38678	&	HIP 27288	&	A2	&	21.5	&	IRAS/MIPS	&	220	&	1.3E-04	&	6	&	N	&	100	&	\citealt{Rhee:2007}	\\
$\beta$ Pic	&	HIP 27321	&	A3	&	19.3	&	IRAS/MIPS	&	110	&	2.6E-03	&	19	&	N	&	12	&	\citealt{Rhee:2007}	\\
HD 40136	&	HIP 28103	&	F1	&	15	&	IRAS/MIPS	&	185	&	2.0E-05	&	6	&	N	&	300	&	\citealt{Rhee:2007}	\\
HD 48682	&	HIP 32480	&	G0	&	16.5	&	IRAS/MIPS	&	60	&	8.9E-05	&	29	&	N	&	600	&	\citealt{Rhee:2007}	\\
HD 50571	&	HIP 32775	&	F7	&	33.2	&	IRAS/MIPS	&	45	&	1.6E-04	&	68	&	N	&	300	&	\citealt{Rhee:2007}	\\
HD 53143	&	HIP 33690	&	K0	&	18.4	&	IRAS/MIPS	&	80	&	2.0E-04	&	9	&	N	&	300	&	\citealt{Rhee:2007}	\\
HD 54341	&	HIP 34276	&	A0	&	92.9	&	IRAS	&	85	&	2.0E-04	&	51	&	N	&	10	&	\citealt{Rhee:2007}	\\
HD 56986	&	HIP 35550	&	F0	&	18	&	IRAS/MIPS	&	60	&	8.9E-06	&	71	&	Y	&	400	&	\citealt{Rhee:2007}	\\
HD 61005	&	HIP 36948	&	G3	&	34.5	&	IRAS/MIPS	&	60	&	2.6E-03	&	16	&	N	&	100	&	\citealt{Rhee:2007}	\\
HD 67523	&	HIP 39757	&	F2	&	19.2	&	IRAS	&	85	&	5.4E-06	&	50	&	N	&	2000	&	\citealt{Rhee:2007}	\\
HD 70313	&	HIP 41152	&	A3	&	51.4	&	IRAS/MIPS	&	80	&	5.2E-05	&	56	&	N	&	200	&	\citealt{Rhee:2007}	\\
HD 71155	&	HIP 41307	&	A0	&	38.3	&	IRAS/MIPS	&	130	&	4.1E-05	&	29	&	N	&	100	&	\citealt{Rhee:2007}	\\
HD 73752	&	HIP 42430	&	G3	&	19.9	&	IRAS	&	80	&	3.2E-05	&	21	&	Y	&	600	&	\citealt{Rhee:2007}	\\
HD 76582	&	HIP 44001	&	F0	&	49.3	&	IRAS	&	85	&	2.2E-04	&	35	&	N	&	300	&	\citealt{Rhee:2007}	\\
HD 84870	&	HIP 48164	&	A3	&	89.5	&	IRAS	&	85	&	5.5E-04	&	32	&	N	&	100	&	\citealt{Rhee:2007}	\\
HD 85672	&	HIP 48541	&	A0	&	93.1	&	IRAS	&	85	&	4.8E-04	&	32	&	N	&	30	&	\citealt{Rhee:2007}	\\
HD 91375	&	HIP 51438	&	A2	&	79.4	&	IRAS	&	85	&	2.4E-05	&	99	&	N	&	400	&	\citealt{Rhee:2007}	\\
HD 91312	&	HIP 51658	&	A7	&	34.3	&	IRAS	&	40	&	1.1E-04	&	179	&	N$^{b}$	&	200	&	\citealt{Rhee:2007}	\\
HD 92945	&	HIP 52462	&	K1	&	21.6	&	IRAS/MIPS	&	45	&	6.7E-04	&	23	&	N	&	100	&	\citealt{Rhee:2007}	\\
HD 95418	&	HIP 53910	&	A1	&	24.3	&	IRAS	&	120	&	1.2E-05	&	45	&	N	&	500	&	\citealt{Rhee:2007}	\\
HD 99945	&	HIP 56253	&	A2	&	59.8	&	IRAS	&	85	&	1.0E-04	&	37	&	N	&	300	&	\citealt{Rhee:2007}	\\
HD 102647	&	HIP 57632	&	A3	&	11.1	&	IRAS/MIPS	&	160	&	4.3E-05	&	11	&	Y	&	50	&	\citealt{Rhee:2007}	\\
HD 107146	&	HIP 60074	&	G2	&	28.5	&	IRAS/MIPS	&	55	&	9.5E-04	&	29	&	N	&	100	&	\citealt{Rhee:2007}	\\
$\eta$ Crv	&	HIP 61174	&	F2	&	18.2	&	IRAS/MIPS	&	180	&	1.2E-04	&	5	&	N	&	300	&	\citealt{Rhee:2007}	\\
HD 110058	&	HIP 61782	&	A0	&	99.9	&	IRAS/IRS	&	130	&	2.5E-03	&	11	&	N	&	10	&	\citealt{Rhee:2007}	\\
HD 110411	&	HIP 61960	&	A0	&	36.9	&	IRAS/MIPS	&	85	&	6.2E-05	&	38	&	N	&	100	&	\citealt{Rhee:2007}	\\
HD 112429	&	HIP 63076	&	F0	&	29	&	MIPS	&	100	&	2.4E-05	&	20	&	N	&	50	&	\citealt{Chen:2005}	\\
HD 113337	&	HIP 63584	&	F6	&	37.4	&	IRAS	&	100	&	1.0E-04	&	18	&	Y	&	50	&	\citealt{Rhee:2007}	\\
HD 115116	&	HIP 64921	&	A7	&	85.4	&	IRAS	&	80	&	3.4E-04	&	39	&	N	&	100	&	\citealt{Rhee:2007}	\\
HD 119124	&	HIP 66704	&	F7	&	25	&	MIPS$^a$	&	55	&	4.0E-05	&	32	&	Y	&	200	&	\citealt{Chen:2005}	\\
HD 121384	&	HIP 68101	&	G8	&	38.1	&	IRAS	&	45	&	2.5E-04	&	91	&	N	&	3000	&	\citealt{Rhee:2007}	\\
HD 122652	&	HIP 68593	&	F8	&	37.2	&	IRAS/MIPS	&	60	&	1.4E-04	&	28	&	N	&	300	&	\citealt{Rhee:2007}	\\
HD 124718	&	HIP 69682	&	G5	&	61.3	&	IRAS	&	85	&	2.1E-03	&	10	&	N	&	500	&	\citealt{Rhee:2007}	\\
HD 125162	&	HIP 69732	&	A0	&	29.8	&	IRAS/MIPS	&	100	&	5.2E-05	&	32	&	N	&	200	&	\citealt{Rhee:2007}	\\
HD 125473	&	HIP 70090	&	A0	&	75.8	&	IRAS	&	120	&	2.1E-05	&	64	&	Y	&	300	&	\citealt{Rhee:2007}	\\
HD 126265	&	HIP 70344	&	G2	&	70.1	&	IRAS	&	85	&	3.9E-04	&	26	&	N	&	500	&	\citealt{Rhee:2007}	\\
HD 127821	&	HIP 70952	&	F4	&	31.7	&	IRAS	&	50	&	2.6E-04	&	55	&	N	&	200	&	\citealt{Rhee:2007}	\\
HD 127762	&	HIP 71075	&	A7	&	26.1	&	IRAS	&	55	&	1.0E-05	&	151	&	Y	&	1000	&	\citealt{Rhee:2007}	\\
$\sigma$ Boo	&	HIP 71284	&	F3	&	15.5	&	IRAS/MIPS/ISO	&	40	&	4.9E-06	&	88	&	N	&	1000	&	\citealt{Rhee:2007}	\\
HD 135502	&	HIP 74596	&	A2	&	69.4	&	IRAS	&	65	&	3.3E-05	&	123	&	N	&	200	&	\citealt{Rhee:2007}	\\
HD 135382	&	HIP 74946	&	A1	&	56	&	IRAS	&	50	&	9.3E-06	&	481	&	N	&	700	&	\citealt{Rhee:2007}	\\
HD 138749	&	HIP 76127	&	B6	&	95.3	&	IRAS	&	75	&	2.0E-05	&	171	&	Y	&	200	&	\citealt{Rhee:2007}	\\
HD 139006	&	HIP 76267	&	A0	&	22.9	&	IRAS/MIPS	&	190	&	2.4E-05	&	17	&	Y	&	500	&	\citealt{Rhee:2007}	\\
HD 139590	&	HIP 76635	&	G0	&	55.1	&	IRAS	&	85	&	3.9E-04	&	17	&	N	&	5000	&	\citealt{Rhee:2007}	\\
HD 139664	&	HIP 76829	&	F5	&	17.5	&	IRAS/MIPS	&	75	&	1.2E-04	&	25	&	N	&	200	&	\citealt{Rhee:2007}	\\
HD 143894	&	HIP 78554	&	A3	&	54.3	&	IRAS	&	45	&	4.6E-05	&	211	&	N	&	300	&	\citealt{Rhee:2007}	\\
HD 149630	&	HIP 81126	&	B9	&	92.7	&	IRAS	&	80	&	3.0E-05	&	157	&	Y	&	700	&	\citealt{Rhee:2007}	\\
HD 150378	&	HIP 81641	&	A1	&	92.9	&	IRAS	&	95	&	1.2E-04	&	57	&	Y	&	200	&	\citealt{Rhee:2007}	\\
HD 151044	&	HIP 81800	&	F8	&	29.4	&	IRAS/MIPS/ISO	&	55	&	8.3E-05	&	35	&	N	&	500	&	\citealt{Rhee:2007}	\\
HD 154145	&	HIP 83480	&	A2	&	94.9	&	IRAS	&	85	&	4.3E-04	&	45	&	N	&	300	&	\citealt{Rhee:2007}	\\
HD 157728	&	HIP 85157	&	F0	&	42.8	&	IRAS	&	90	&	2.7E-04	&	30	&	N	&	100	&	\citealt{Rhee:2007}	\\
HD 158352	&	HIP 85537	&	A8	&	63.1	&	IRAS/MIPS	&	70	&	6.8E-05	&	85	&	N	&	600	&	\citealt{Rhee:2007}	\\
HD 161868	&	HIP 87108	&	A0	&	29.1	&	IRAS/MIPS	&	85	&	7.8E-05	&	54	&	N	&	200	&	\citealt{Rhee:2007}	\\
HD 162917	&	HIP 87558	&	F4	&	31.4	&	IRAS	&	85	&	2.5E-04	&	20	&	N	&	400	&	\citealt{Rhee:2007}	\\
HD 164249	&	HIP 88399	&	F5	&	46.9	&	IRAS/MIPS/ISO	&	70	&	1.0E-03	&	27	&	N	&	12	&	\citealt{Rhee:2007}	\\
HD 169022	&	HIP 90185	&	B9	&	44.3	&	IRAS	&	100	&	4.5E-06	&	155	&	Y	&	300	&	\citealt{Rhee:2007}	\\
HD 170773	&	HIP 90936	&	F5	&	36.1	&	IRAS/MIPS/ISO	&	50	&	4.6E-04	&	61	&	N	&	200	&	\citealt{Rhee:2007}	\\
Vega	&	HIP 91262	&	A0	&	7.8	&	IRAS/MIPS	&	80	&	2.1E-05	&	93	&	N	&	220	&	\citealt{Rhee:2007}	\\
HD 172555	&	HIP 92024	&	A7	&	29.2	&	IRAS/MIPS	&	320	&	8.1E-04	&	2	&	Y	&	12	&	\citealt{Rhee:2007}	\\
PZ Tel	&	HIP 92680	&	K0	&	49.7	&	MIPS$^a$	&	85	&	9.4E-05	&	11	&	Y	&	12	&	\citealt{Rebull:2008}	\\
HD 176638	&	HIP 93542	&	A0	&	56.3	&	IRAS	&	120	&	9.7E-05	&	34	&	N	&	200	&	\citealt{Rhee:2007}	\\
$\eta$ Tel	&	HIP 95261	&	A0	&	47.7	&	IRAS/MIPS/ISO	&	150	&	2.1E-04	&	15	&	Y	&	12	&	\citealt{Rhee:2007}	\\
HD 181327	&	HIP 95270	&	F5	&	50.6	&	IRAS/MIPS	&	75	&	3.5E-03	&	25	&	N	&	12	&	\citealt{Rhee:2007}	\\
HD 182681	&	HIP 95619	&	B8	&	69.1	&	IRAS	&	85	&	2.0E-04	&	55	&	N	&	50	&	\citealt{Rhee:2007}	\\
HD 191089	&	HIP 99273	&	F5	&	53.5	&	IRAS/MIPS	&	95	&	1.4E-03	&	15	&	N	&	30	&	\citealt{Rhee:2007}	\\
HD 191692	&	HIP 99473	&	B9	&	88	&	IRAS	&	85	&	6.6E-06	&	213	&	Y	&	500	&	\citealt{Rhee:2007}	\\
HD 195627	&	HIP 101612	&	F1	&	27.6	&	IRAS/MIPS	&	65	&	1.1E-04	&	51	&	N	&	200	&	\citealt{Rhee:2007}	\\
HD 196544	&	HIP 101800	&	A2	&	54.3	&	IRAS/MIPS	&	100	&	3.9E-05	&	31	&	Y	&	30	&	\citealt{Rhee:2007}	\\
AU Mic	&	HIP 102409	&	M1	&	9.9	&	IRAS/MIPS	&	50	&	3.6E-04	&	9	&	N	&	12	&	\citealt{Rhee:2007}	\\
HD 205674	&	HIP 106741	&	F3	&	52.6	&	IRAS	&	85	&	4.0E-04	&	20	&	N	&	300	&	\citealt{Rhee:2007}	\\
HD 205536	&	HIP 107022	&	G8	&	22.1	&	IRAS	&	80	&	2.9E-04	&	10	&	N	&	500	&	\citealt{Rhee:2007}	\\
HD 206893	&	HIP 107412	&	F5	&	38.9	&	IRAS/MIPS/ISO	&	55	&	2.7E-04	&	41	&	N	&	200	&	\citealt{Rhee:2007}	\\
HD 207129	&	HIP 107649	&	G2	&	15.6	&	IRAS/MIPS/ISO	&	55	&	1.2E-04	&	27	&	Y	&	600	&	\citealt{Rhee:2007}	\\
HD 209253	&	HIP 108809	&	F6	&	30.1	&	IRAS/MIPS/ISO	&	75	&	7.3E-05	&	18	&	N	&	200	&	\citealt{Rhee:2007}	\\
%HD 211336	&	HIP 109857	&	F0	&	25.7	&	IRAS	&	65	&	1.6E-04	&	62	&	N$^{c}$	&	600	&	\citealt{Rhee:2007}	\\
HD 213617	&	HIP 111278	&	F1	&	52.9	&	IRAS/MIPS	&	55	&	9.4E-05	&	69	&	N	&	600	&	\citealt{Rhee:2007}	\\
Fomalhaut	&	HIP 113368	&	A3	&	7.7	&	IRAS/MIPS	&	65	&	8.0E-05	&	73	&	Y	&	220	&	\citealt{Rhee:2007}	\\
HD 218396	&	HIP 114189	&	A5	&	39.9	&	IRAS/ISO	&	50	&	2.3E-04	&	77	&	N	&	30	&	\citealt{Rhee:2007}	\\
HD 221853	&	HIP 116431	&	F0	&	71.2	&	IRAS/MIPS/ISO	&	85	&	7.4E-04	&	26	&	N	&	100	&	\citealt{Rhee:2007}	\\
\enddata
\tablecomments{ \,
List of all debris disks stars in our sample. $R_\text{dust}$ is the disk radius for blackbody grains (see Section~\ref{dust_T}).
 \\$^a$--- $T_\text{dust}$, $L_{IR}/L_*$, and $R_\text{dust}$ have been modified from those given in the cited references so that the peak of the blackbody fit lies at the MIPS 70\micron\, detection (see Section~\ref{dust_T} for more details).
 \\$^b$--- While HIP~51658 is listed in the 7th spectroscopic binary orbit catalog with a period of 292.56~days \citep{Batten:1978}, the quality flag is set to the lowest value indicating the binary nature is in question. More recent catalogs (see \citealt{Pourbaix:2004}) no longer list this system.
% \\$^c$--- Although, \citet{Eggleton:2008} list HIP~109857 as a probable binary system, they provide no separation or period. The Washington Double Star catalog lists 3 candidate companions \citep{Mason:2001}, however, all can be ruled out as their proper motions and photometric distances differ from the primary by a significant amount.
 }
\label{tab:debris_sample}
\end{deluxetable}

%% file: debris_binaries_singlespace.tex
\begin{deluxetable}{llcc|cc|cc|p{5cm}}
%\begin{deluxetable}{llcc|ccc|ccc|l}
\tablecolumns{9}
\tabletypesize{\scriptsize}
\rotate
\tablewidth{0pc}
\tablecaption{Multiples in Debris Disk Sample}

\tablehead{
\colhead{Name} & \colhead{Spectral} & \colhead{Dist.} & \colhead{$R_\text{dust}$} & 
\colhead{S1} & \colhead{S2} & \colhead{P1} & \colhead{P2}  & \colhead{References}    \\
\colhead{} & \colhead{Type} & \colhead{(pc)} & \colhead{(AU)} & \colhead{(AU)} & \colhead{(AU)} & \colhead{(days)} & \colhead{(days)} & \colhead{} }

\startdata
HIP 1185	&	A7	&	88.3	&	173	&	626.9	&	\nodata	&	4.27E+06	&	\nodata	&	this work	\\ \hline
HIP 6686	&	A5	&	30.5	&	88	&	2.1	&	\nodata	&	7.59E+02	&	\nodata	&	\citealt{Eggleton:2008,Samus:2009}	\\ \hline
HIP 8920	&	G0	&	91.9	&	0.85	&	0.047	&	\nodata	&	3.45E+00	&	\nodata	&	\citealt{Weinberger:2008}	\\ \hline
HIP 10679	&	G2	&	34	&	20	&	471.9	&	\nodata	&	3.06E+06	&	\nodata	&	\citealt{ESA:1997}	\\ \hline
HIP 11437$^{a}$	&	K8	&	42.3	&	10	&	1015.2	&	\nodata	&	1.18E+07	&	\nodata	&	\citealt{Zuckerman:2004,Mason:2011}	\\ \hline
HIP 14576	&	B8	&	28.5	&	13	&	0.07	&	3.0	&	2.87E+00	&	6.80E+02	&	\citealt{Pourbaix:2004, Eggleton:2008, Tokovinin:1997}	\\ \hline
HIP 15197	&	A5	&	36.8	&	31	&	0.17	&	\nodata	&	1.79E+01	&	\nodata	&	\citealt{Pourbaix:2004, Eggleton:2008, Trilling:2007}	\\ \hline
HIP 25486	&	F7	&	26.8	&	18.9	&	0.021	&	\nodata	&	9.00E-01	&	\nodata	&	\citealt{Eker:2008}	\\ \hline
HIP 27072	&	F7	&	9	&	15	&	866.3	&	\nodata	&	8.17E+06	&	\nodata	&	\citealt{Gould:2004, Eggleton:2008}	\\ \hline
HIP 35550	&	F0	&	18	&	71	&	3.9	&	125.6	&	2.24E+03	&	4.37E+05	&	\citealt{Pourbaix:2004, Eggleton:2008,Tokovinin:1997}	\\ \hline
HIP 42430	&	G3	&	19.9	&	21	&	34.0	&	\nodata	&	4.49E+04	&	\nodata	&	\citealt{ESA:1997, Eggleton:2008, Raghavan:2010}	\\ \hline
HIP 57632	&	A3	&	11.1	&	11	&	440.7	&	\nodata	&	2.39E+06	&	\nodata	&	\citealt{Eggleton:2008}	\\ \hline
HIP 63584	&	F6	&	37.4	&	18	&	4450.6	&	\nodata	&	9.51E+07	&	\nodata	&	\citealt{Reid:2007,Mason:2001}	\\ \hline
HIP 66704	&	F7	&	25	&	31.9	&	440.0	&	\nodata	&	2.96E+06	&	\nodata	&	\citealt{Duquennoy:1991}	\\ \hline
HIP 70090	&	A0	&	75.8	&	64	&	0.33	&	\nodata	&	3.88E+01	&	\nodata	&	\citealt{Eggleton:2008}	\\ \hline
HIP 71075	&	A7	&	26.1	&	151	&	1.8	&	\nodata	&	6.72E+02	&	\nodata	&	\citealt{Eggleton:2008}	\\ \hline
HIP 76127	&	B6	&	95.3	&	171	&	66.3	&	\nodata	&	8.06E+04	&	\nodata	&	\citealt{ESA:1997, Alzner:1998}	\\ \hline
HIP 76267	&	A0	&	22.9	&	17	&	0.19	&	\nodata	&	1.74E+01	&	\nodata	&	\citealt{Pourbaix:2004, Eggleton:2008}	\\ \hline
HIP 81126$^{b}$	&	B9	&	92.7	&	157	&	10.2	&	\nodata	&	5.56E+03	&	\nodata	&	Sixth Catalog of Orbits of Visual Binary Stars	\\ \hline
HIP 81641	&	A1	&	92.9	&	57	&	13.7	&	6477.0	&	9.12E+03	&	7.24E+07	&	\citealt{Tokovinin:1997, Eggleton:2008}	\\ \hline
HIP 90185	&	B9	&	44.3	&	155	&	105.9	&	\nodata	&	2.13E+05	&	\nodata	&	\citealt{Eggleton:2008}	\\ \hline
HIP 92024	&	A7	&	29.2	&	2	&	1927.2	&	\nodata	&	2.30E+07	&	\nodata	&	\citealt{Eggleton:2008}	\\ \hline
\tablebreak
HIP 92680	&	K0	&	49.7	&	11.3	&	19.9	&	\nodata	&	2.74E+04	&	\nodata	&	\citealt{Biller:2010}	\\ \hline
HIP 95261	&	A0	&	47.7	&	15	&	198.9	&	\nodata	&	6.91E+05	&	\nodata	&	\citealt{Eggleton:2008}	\\ \hline
HIP 99473	&	B9	&	88	&	213	&	0.28	&	\nodata	&	1.71E+01	&	\nodata	&	\citealt{Shatskii:1998, Pourbaix:2000, Pourbaix:2004, Eggleton:2008}	\\ \hline
HIP 101800	&	A2	&	54.3	&	31	&	0.15	&	\nodata	&	1.10E+01	&	\nodata	&	\citealt{Pourbaix:2004, Eggleton:2008}	\\ \hline
HIP 107649	&	G2	&	15.6	&	27	&	858.0	&	\nodata	&	8.38E+06	&	\nodata	&	\citealt{Eggleton:2008}	\\ \hline
HIP 113368	&	A3	&	7.7	&	73	&	54380	&	\nodata	&	3.28E+09	&	\nodata	&	\citealt{Barrado:1997}	\\ \hline
\enddata
\tablecomments{ \,
List of multiples in our debris disk sample. Columns S1 and S2 denote the binary/multiple separation and likewise P1 and P2 denote the period of the orbit.
\\$^a$--- While HIP~11437 is not explicitly labeled as a binary in the $\beta$~Pic moving group \citep{Zuckerman:2004,Torres:2008}, the kinematics, distance, and close proximity of BD+30~397B suggest these two stars are a physically bound $\sim$1000~AU binary system.
\\$^b$--- Two periods are commonly listed for HIP~81126: $P = 15.2$ or $7.48$ years ($a = 0.11$ or $0.074\arcsec$; \citealt{Mason:2001}).
While \citet{Eggleton:2008} adopt the 7.48-year period, we chose to use the 15.2-year period as this is fit well in the Sixth Orbit Catalog of Visual Binary Stars with recent measurements \citep{Brendley:2007} and is the period adopted in other recent papers \citep{Cvetkovic:2010,Mason:2009}.
 }
\label{tab:multiples}
\end{deluxetable}

%% file: resolved_disks.tex
\begin{deluxetable}{lllcccr}
\tablecolumns{9}
\tabletypesize{\scriptsize}
%\rotate
\tablewidth{0pc}
\tablecaption{Resolved Debris Disks}

\tablehead{
\colhead{Name} & \colhead{$R_\text{res}$} & \colhead{$\lambda$} & \colhead{Type} & \colhead{$T_\text{dust}$} & \colhead{$R_\text{BB}$} & \colhead{Refs} \\
\colhead{} & \colhead{(AU)} & \colhead{(\micron)} & \colhead{} & \colhead{(K)} & \colhead{(AU)} & \colhead{} }

\startdata
HD 139664	&	83	&	0.6	&	scattered	&	80, 50	&	21, 61	&	\citealt{Kalas:2006,Chen:2009}	\\ % r_scat= 83-109 , peak at 83
%HD 15115		&	430\tablenotemark{a}	&	0.61	&	scattered	&	65	&	35	&	\citealt{Kalas:2007a,Rhee:2007}	\\
%HD 61005		&	210\tablenotemark{b}	&	1.1	&	scattered	&	58	&	17	&	\citealt{Hines:2007,Hillenbrand:2008}	\\
%HD~53143	&	82\tablenotemark{c}		&	0.6	&	scattered	&	80	&	9	&	\citealt{Kalas:2006,Rhee:2007}	\\
%AU Mic		&	130\tablenotemark{d}	&	0.6	&	scattered	&	50	&	9	&	\citealt{Kalas:2004,Rhee:2007}	\\
HD 15115		&	430	&	0.61	&	scattered	&	65	&	35	&	\citealt{Kalas:2007a,Rhee:2007}	\\
HD 61005		&	210	&	1.1	&	scattered	&	58	&	17	&	\citealt{Hines:2007,Hillenbrand:2008}	\\
HD~53143	&	82		&	0.6	&	scattered	&	80	&	9	&	\citealt{Kalas:2006,Rhee:2007}	\\
AU Mic		&	130	&	0.6	&	scattered	&	50	&	9	&	\citealt{Kalas:2004,Rhee:2007}	\\
HD 207129	&	163	&	0.6	&	scattered	&	55	&	27	&	\citealt{Krist:2010,Rhee:2007}	\\
HD 181327	&	86	&	1.1	&	scattered	&	75	&	25	&	\citealt{Schneider:2006,Rhee:2007}	\\
HD 15745		&	300	&	0.59	&	scattered	&	85	&	22	&	\citealt{Kalas:2007b,Rhee:2007}	\\
HR 4796A		&	70	&	1.1	&	scattered	&	110	&	30	&	\citealt{Schneider:1999,Rhee:2007}	\\
Fomalhaut	&	141	&	0.6	&	scattered	&	65	&	73	&	\citealt{Kalas:2005,Rhee:2007}	\\
HD 141569A	&	185	&	1.1	&	scattered	&	110	&	24	&	\citealt{Weinberger:1999,Rhee:2007}	\\
HD~92945	&	54	&	0.6	&	scattered	&	45	&	23	&	\citealt{Golimowski:2011,Rhee:2007}	\\
\hline
Vega		&	85	&	70, 160	&	thermal	&	80	&	93	&	\citealt{Sibthorpe:2010,Rhee:2007}	\\
HD 107146	&	97	&	1300	&	thermal	&	55	&	29	&	\citealt{Corder:2009,Rhee:2007}	\\
HD 32297		&	115	&	12--19	&	thermal	&	85	&	28	&	\citealt{Moerchen:2007,Rhee:2007}	\\
HD 10647		&	85	&	70	&	thermal	&	65	&	22	&	\citealt{Liseau:2010,Rhee:2007}	\\
HR 8799		&	200	&	70	&	thermal	&	150, 45	&	9, 95	&	\citealt{Su:2009}	\\
HD 191089	&	59	&	18.2	&	thermal	&	95	&	15	&	\citealt{Churcher:2011,Rhee:2007}	\\
$\eta$ Tel		&	24	&	18.3	&	thermal	&	150	&	15	&	\citealt{Smith:2009,Rhee:2007}	\\ %HD 181296
$\epsilon$ Eri	&	60	&	850	&	thermal	&	40	&	27	&	\citealt{Greaves:1998,Rhee:2007}	\\
$\eta$ Crv	&	145	&	100	&	thermal	&	354, 31	&	174, 1.4	&	\citealt{Matthews:2010}	\\
$\beta$ UMa	&	47	&	100	&	thermal	&	109	&	51	&	\citealt{Matthews:2010}	\\
$\beta$ Leo	&	39	&	100	&	thermal	&	112	&	23	&	\citealt{Matthews:2010}	\\
HD 98800B	&	13	&	880	&	thermal	&	160	&	2.2	&	\citealt{Andrews:2010,Low:2005}	\\
\enddata

%\tablenotetext{a}{HD~15115's disk is asymmetric: 315~AU and $>550$~AU; we quote the average.}
%\tablenotetext{b}{For HD~61005, we quote the approximate maximum radial extent (210~AU) of the unusual disk structure seen in the system, see \citet{Hines:2007} for details.}
%\tablenotetext{c}{We quote an average dust radius of 82~AU for HD~53143, which has constraints of $>55$ and $<110$~AU for the dust in the system based on sensitivity limits.}
%\tablenotetext{d}{For AU~Mic, we quote the average of the outer radius (201~AU) and inner limit (50~AU) presented in \citet{Kalas:2004}, but subsequent measurements have demonstrated the disk extends farther in to $\sim$8~AU (see \citealt{Fitzgerald:2007,Krist:2005}).}

\tablecomments{
Disk radii ($R_{res}$) for resolved systems. 
This table is not meant to represent the most descriptive values for the system (for example, some disks are clearly asymmetric); we refer the reader to the individual references for details.
For systems resolved in both thermal and scattered emission, we list only the thermally resolved radius. 
%Resolved radii come from the respective references; we average the range provided (many are rings or belts) or adopt the location of peak emission. 
Resolved and blackbody-fit radii are compared in Figure~\ref{fig:resolved}.
%A handful of known resolved systems are not listed, such as $\beta$~Pic, $\tau$~Ceti, $\gamma$~Oph, and $\zeta^2$~Ret as either too little information is available (or in $\beta$~Pic's case, too much; for example, see the note on AU~Mic) and thus meaningful determinations of a characteristic resolved radius are not possible. We note that HD~141569A and HD~98800B are classified as transition disks.
 }
\label{tab:resolved}
\end{deluxetable}